\documentclass[aps,pra,groupedaddress,twocolumn]{revtex4-1}

\usepackage{graphicx}
\usepackage{dcolumn}
\usepackage{bm}
\usepackage{amsmath}
\usepackage{amssymb}
\usepackage{latexsym}
\usepackage{epsfig}
\usepackage{amsbsy}
\usepackage{array}
\usepackage{amssymb}
\usepackage{setspace}
\usepackage{bm}
\usepackage{epstopdf}
\usepackage{subfigure}
\usepackage[colorlinks, linkcolor=blue, anchorcolor=blue, citecolor=blue, citecolor=blue,unicode=true]{hyperref}

\def\sint{\ifmmode{- \!\!\!\!\!\! \int}
    \else{\hbox{$- \!\!\!\! \int \ $}}\fi}

\begin{document}

\title{Creation of superposition of arbitrary states encoded in two three-dimensional cavities}
\author{Tong Liu$^1$, Yang Zhang$^2$, Bao-qing Guo$^1$}
\author{Chang-shui Yu$^1$}
\email{ycs@dlut.edu.cn}
\author{Wei-ning Zhang$^1$}

\address{$^1$School of Physics, Dalian University of Technology, Dalian 116024, China}
\address{$^2$Institute of Theoretical Physics, Shanxi Datong University, Datong 037009, China}

\date{\today}

\begin{abstract}
The principle of superposition is a key ingredient for quantum mechanics. A recent work [M. Oszmaniec \textit{et al.}, Phys. Rev. Lett. 116, 110403 (2016)] has shown that a quantum adder that deterministically generates a superposition of two unknown states is forbidden. Here we propose a probabilistic approach for creating a superposition state of two arbitrary states
encoded in two three-dimensional cavities. Our implementation is based
on a three-level superconducting transmon qubit dispersively
coupled to two cavities.
Numerical
simulations show that high-fidelity generation of the superposition of two coherent states is feasible with current circuit QED technology.
Our method also works for other physical systems such as other types of superconducting qubits, natural atoms, quantum dots, and nitrogen-vacancy (NV) centers.
\end{abstract}

\maketitle

\section{Introduction}
\label{sec1}
The superpositions of quantum states is at the heart of
the basic postulates and theorems of quantum mechanics. Quantum superposition studies the application of quantum theory or phenomena, that is different from the classical world,
leads to many other intriguing quantum
phenomena such as quantum entanglement~\cite{09entanglement} and quantum coherence~\cite{14coherence,17coherence}.
It is a vital
physical resource and has many important applications in quantum information processing (QIP) and quantum computation such as quantum algorithms~\cite{97Shor,96Grover}, quantum metrology~\cite{11Giovannetti},
and quantum cryptography~\cite{02cryptography}.

A quantum adder is a quantum machine adding two arbitrary unknown quantum
states of two different systems onto a
single system~\cite{15adder,16adder}.
How to generate a superposition of two arbitrary states has
recently aroused great interest in the field of quantum optics and quantum information. For example, Refs.~\cite{15adder,16adder} have proved that it is impossible to generate a superposition of two unknown states, but Ref.~\cite{16adder}
proposed a method to
probabilistically 
creating the superposition of two known pure states
with the fixed overlaps.
Ref.~\cite{17adder} has shown that
superpositions of orthogonal
qubit states can be produced with unit probability, and
Ref.~\cite{17adder2} has demonstrated that the
state transfer can be protected via an
approximate quantum adder.
Recently, the probabilistic creation of superposition of two unknown quantum states has been demonstrated experimentally
in linear optics~\cite{16huang} and nuclear magnetic resonance (NMR)~\cite{17Laflamme}.

Circuit quantum electrodynamics (QED) consisting of superconducting qubits and microwave
cavities are now moving toward
multiple superconducting qubits, multiple three-dimensional (3D) cavities with greatly enhanced coherence time, making them particularly appealing for large-scale quantum computing \cite{11you,13Devoret,17liuy}.
For example, a 3D microwave cavity with the photon lifetime
up to 2 $s$~\cite{18Romanenko} and a transmon with a coherence time $ \backsim $ 0.1 $ms$ \cite{12Rigetti} have been recently reported in 3D circuit QED. Hence, 3D cavities are good memory elements, which can have coherence time at least
four orders of magnitude longer than the transmons.
By encoding quantum information in microwave cavities,
many schemes
have been proposed for synthesizing Bell states \cite{08Mariantoni},
NOON states~\cite{njpMerkel,10Strauch,15Xiong,16Sharma,16Zhao,17su,18liut}, and entangled coherent states~\cite{12yang,16liu2} of multiple cavities, and realizing cross-Kerr nonlinearity interaction between two cavities~\cite{17liu,18deng}.

Three-dimensional circuit QED has emerged as a
well-established platform for QIP and quantum computation \cite{s6,13Vlastakis,15transmon,16correction,16Wangc,16Rosenblum}, including 
creation of a Schr\"{o}dinger cat state of a microwave cavity \cite{13Vlastakis}, preparation and control of a five-level transmon qudit \cite{15transmon},
demonstration of a quantum error correction \cite{16correction}, 
realization of a two-mode cat state of two
microwave cavities \cite{16Wangc}, and implementation of a controlled-NOT gate between multiphoton qubits encoded
in two cavities \cite{16Rosenblum}.
Considering these advancements in 3D circuit QED, it is quite
meaningful and necessary to implement a quantum adder in
such systems.

In this paper, we present a probabilistic scheme to realize a quantum adder that creates a superposition of two unknown states by using a superconducting transmon qubit dispersively
coupled to two 3D cavities. This circuit architecture
has been experimentally demonstrated recently in \cite{16Wangc}. Our protocol has the following features and advantages: (\romannumeral1)
The superposition of two states are encoded in two three-dimensional cavities which have long coherence time
rather than encoded in qubits. 
(\romannumeral2) The states of cavities can be
in arbitrary states, e.g., discrete-variable states or continuous-variable states.
(\romannumeral3) Due to the interaction between the two cavities is
mediated by the transmon, cavity-induced dissipations are greatly suppressed.
(\romannumeral4) 
Our proposal can also be applied to
other physical systems such as quantum
dot-cavity system \cite{07dot}, natural atom-cavity system \cite{05atom}, superconducting circuits with other types of superconducting qubits (e.g.,  phase qubit \cite{11phase},  Xmon 
qubit \cite{13Xmon}, flux qubit \cite{16flux}), hybrid circuits for two nitrogen-vacancy center ensembles coupled to a single
flux qubit \cite{11zhu}.

The remaining of this paper is  organized as follows.
In Sec.~\ref{sec2}, we review some basic theory of quantum adder that creates the superposition of two arbitrary states.
Our experimental system and Hamiltonian are 
introduced in Sec.~\ref{sec3}. In Sec.~\ref{sec4},
we show the method to implementing a quantum adder in 3D
circuit QED system. In Sec.~\ref{sec5}, we give a brief discussion on the experimental implementation of a quantum adder with  state-of-the-art circuit QED technology. Finally,
Sec. \ref{sec6} gives a brief concluding summary.

\section{BASIC THEORY OF QUANTUM ADDER} \label{sec2}

\begin{figure}[tbp]
\begin{center}
\includegraphics[bb=95 598 445 704, width=8.0 cm, clip]{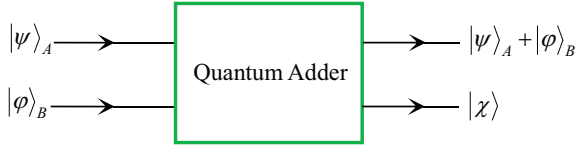} \vspace*{%
-0.08in}
\end{center}
\caption{(color online) A diagram of the quantum adder. Here,
$\vert\psi\rangle_a$ and $\vert \varphi
\rangle_b$ are the two arbitrary input pure states, while
$\vert\psi\rangle_a+\vert \varphi
\rangle_b$ is the out superposition state with the referential state $\vert\chi\rangle$.}
\label{fig:1}
\end{figure}

We now review some basic theory of quantum adder (Fig.~1) which can generate the superposition of two
pure states. Assume that particles $A$ and $B$ are respectively initially in arbitrary pure states $\vert\psi\rangle_A$ and $\vert \varphi \rangle_B$, and an ancilla particle $T$ is prepared in an arbitrary superposition state $\vert\phi\rangle_T=\alpha_{T}\vert 0\rangle_T+\beta_{T}\vert 1
\rangle_T$ with normalized coefficients $\alpha_{T}$ and $\beta_{T}$.
We first implement a three-qubit controlled-SWAP gate such that the initial state $\vert\psi\rangle_A\vert \varphi \rangle_B\vert\phi\rangle_T$ of three qubits evolves into
\begin{eqnarray}  \label{eq1}
\alpha_{T}|\psi\rangle_A|\varphi\rangle_B|0\rangle_T
+\beta_{T}|\varphi\rangle_A|\psi\rangle_B|1\rangle_T,
\end{eqnarray}
where the qubit $T$ is a control qubit and qubits $A$ and $B$
are two target qubits. Equation~(\ref{eq1}) means that the states of the target qubits are swapped only if
the the control qubit is in the state $|1\rangle_T$ and unchanged otherwise.

Then we make projective measurements on the states $|\pm\rangle_T$ and $|\chi\rangle_B$ of qubits
$T$ and $A$, respectively. Here, $|\pm\rangle_T=(|0\rangle_T \pm|1\rangle_T )/\sqrt{2}$ and $|\chi\rangle_B$ is the referential state which satisfies $\langle \chi |\psi\rangle_B\neq 0$ and $\langle \chi |\varphi\rangle_B\neq 0$. Accordingly, one obtains the following superposition state of qubit $A$
\begin{eqnarray}  \label{eq2}
|\Psi\rangle_A= \frac{1}{N}(\gamma|\psi\rangle
\pm\eta|\varphi\rangle),
\end{eqnarray}
where $\gamma=\alpha_{T}\langle\chi\vert\varphi\rangle_B$, 
$\eta=\beta_{T}\langle\chi\vert\psi\rangle_B$, and the normalization constant $N=\sqrt{(1/2)[\vert\gamma\vert^2+\vert\eta\vert^2\pm2\textrm{Re}(\gamma \eta^*\langle\varphi\vert\psi\rangle)]}$. Here, the sign $``+"$ or $``-"$ of the output state depending on the
measurement $|+\rangle$ or $|-\rangle$ of 
ancilla qubit $T$. It can be seen that the performance of the  above superposition state is possible with prior knowledge of
the overlaps of $\langle\chi\vert\varphi\rangle$ and
$\langle\chi\vert\psi\rangle$.
As shown in above operations, the superposition state of Eq.~(\ref{eq2}) can be produced under the knowledge about the overlaps $\langle\chi\vert\varphi\rangle$ and
$\langle\chi\vert\psi\rangle$. We find that if we vary the coefficients $\alpha_T$ and $\beta_T$, then the arbitrary
superposition state of qubit $A$ is prepared with prior knowledge of
the overlaps.

\section{System and Hamiltonian}\label{sec3}
\begin{figure}[tbp]
\begin{center}
\includegraphics[bb=163 211 457 391, width=8.0 cm, clip]{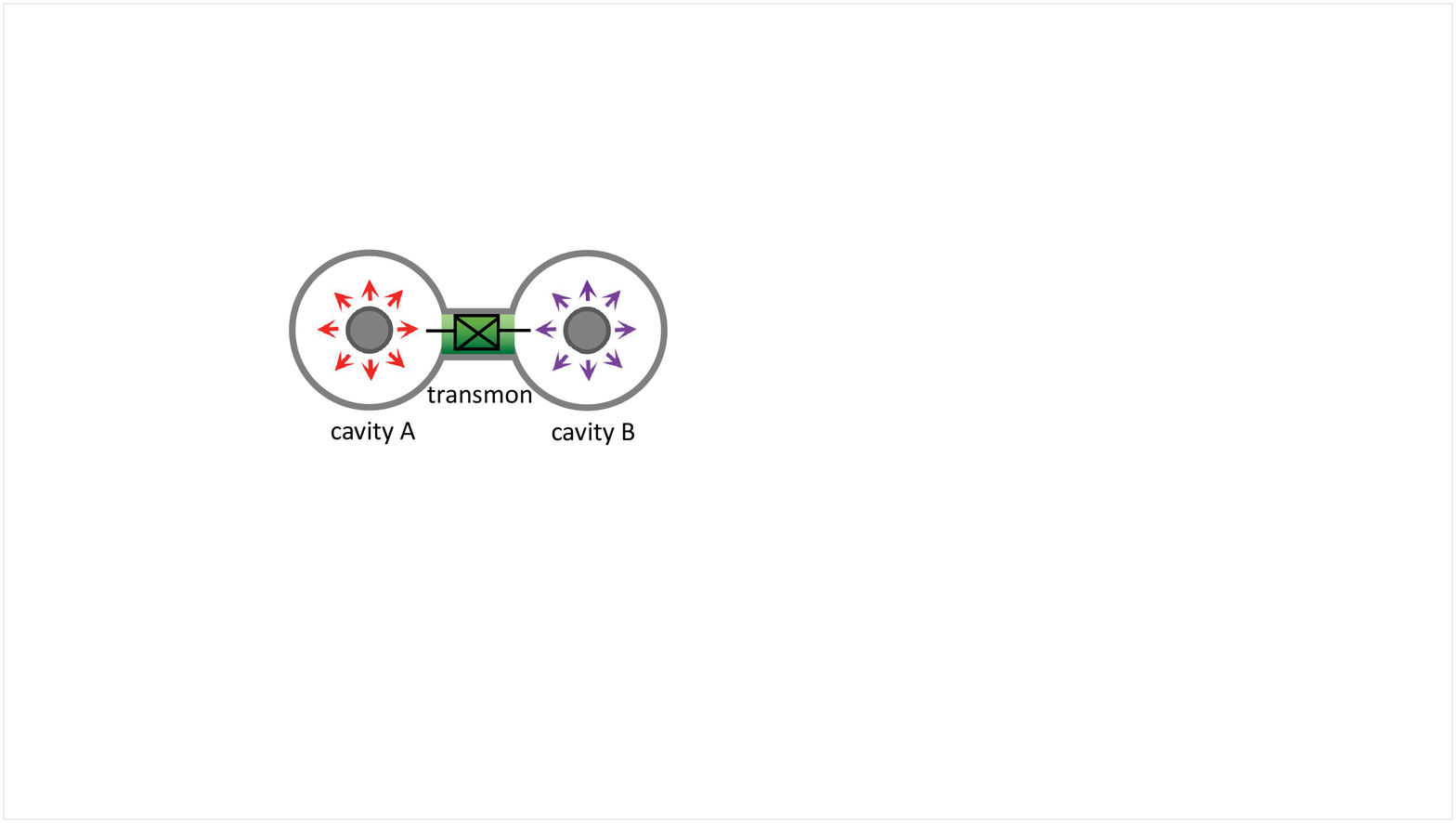}
\vspace*{-0.08in}
\end{center}
\caption{(color online) Schematic of a single transmon qubit dispersively coupled to two three-dimensional microwave cavities $A$ and $B$.}
\label{fig:}
\end{figure}

Superconducting qubits based on Josephson junctions
are mesoscopic element circuits like artificial atoms, with multiple discrete
energy levels whose spacings can be rapidly adjusted by varying external control
parameters \cite{13Xmon,08M. Neeley,09P. J. Leek,13J. D. Strand}.
Typically, a transmon \cite{07trans}  has weakly anharmonic multilevel structure and the transition between non-adjacent levels is forbidden or very weak.  

Motivated by the experimental advances in 3D circuit
QED, we here consider a circuit system consisting of a
transmon qutrit (with three states $\vert g\rangle$, $\vert e\rangle$ and $\vert f\rangle$) capacitively coupled to two separate 3D superconducting cavities as shown in Fig.~2.
The Hamiltonian to describe the microwave cavities
coupled to the transmon reads $H=H_0+H_I$. The free Hamiltonian $H_0$ is given by
\begin{eqnarray} \label{eq3}
H_{0}=\omega_{eg}\vert e\rangle\langle e\vert+(\omega_{eg}+\omega_{fe})\vert f\rangle\langle f\vert+\omega_{A}a^{\dagger}a+\omega_{B}b^{\dagger}b, \nonumber \\
\end{eqnarray}
with $\omega_{eg}$ ($\omega_{fe}$) denoting the $\vert g\rangle\leftrightarrow\vert e\rangle$ ($\vert e\rangle\leftrightarrow\vert f\rangle$) transition frequency of
transmon qubit, $\omega_{A}$ ($\omega_{B}$) denoting the frequency of cavity $A$ ($B$), and $a^{\dagger}$ and $a$ ($b^{\dagger}$ and $b$) representing the creation and annihilation operators for cavity $A$ ($B$).

The coupling Hamiltonian $H_I$ of the whole system is given by
\begin{eqnarray} \label{eq101}
&H_{I}=g_A(a\sigma_{eg}^{+}+a^{\dagger}\sigma_{eg}^{-})+g_B(b\sigma_{eg}^{+}+b^{\dagger}\sigma_{eg}^{-})\notag \\
&+\sqrt{2}g_A(a\sigma_{fe}^{+}+a^{\dagger}\sigma_{fe}^{-})+\sqrt{2}g_B(b\sigma_{fe}^{+}+b^{\dagger}\sigma_{fe}^{-}),
\end{eqnarray}
where  $\sigma_{eg}^{+}=|e\rangle\langle g|$ ($\sigma_{eg}^{-}=|g\rangle\langle e|$), $\sigma_{fe}^{+}=|f\rangle\langle e|$ ($\sigma_{fe}^{-}=|e\rangle\langle f|$) represent the raising (lowering) operators respectively corresponding to the $\vert g\rangle\leftrightarrow\vert e\rangle$ and the $\vert e\rangle\leftrightarrow\vert f\rangle$  transitions in the transmon, $g_A$ ($g_B$)  and $\sqrt{2}g_A$ ($\sqrt{2}g_B$) denote the coupling strengths.
Here we would like to emphasize that both the transitions in the transmon should be considered due to the weak anharmonicity  \cite{07trans}.
However,  the $\vert g\rangle\leftrightarrow\vert f\rangle$ transition of the transmon is a forbidden dipole transition~\cite{07trans}, so the transition $\vert g\rangle\leftrightarrow\vert f\rangle$ can be neglected.
\begin{figure}[tbp]
\begin{center}
\includegraphics[bb=364 491 583 713, width=8 cm, clip]{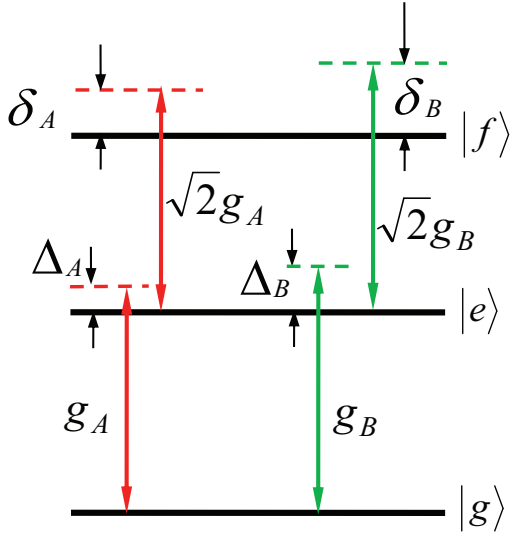}
\vspace*{-0.08in}
\end{center}
\caption{ (color online) Schematic diagram of transmon-cavity interaction.
Cavity $j$ is far-off resonant with the $
\left\vert g\right\rangle \leftrightarrow \left\vert e\right\rangle $ ($\left\vert e\right\rangle \leftrightarrow \left\vert
f\right\rangle $)
transition of transmon qubit with coupling strength $g_{j}$ ($\sqrt{2}g_j$)
and detuning $\Delta _{j}$ ($\delta_j$). Here, $\Delta_{j}=\omega_{eg}-\omega_{j}$ and $\delta _{j}=\omega_{fe}-\omega_{j}$ ($j=A,B$).}
\label{fig:}
\end{figure}

To proceed, let's turn to the the interaction picture with respect to
$H_{0}=\omega_{eg}\vert e\rangle\langle e\vert+(\omega_{eg}+\omega_{fe})\vert f\rangle\langle f\vert+\omega_{A}a^{\dagger}a+\omega_{B}b^{\dagger}b$. Then the
Hamiltonian can be given by
\begin{eqnarray} \label{eq102}
H_{I}&=&g_A(a\sigma_{eg}^{+}e^{i\Delta_A t}+a^{\dagger}\sigma_{eg}^{-}e^{-i\Delta_A t})\notag \\
&+&g_B(b\sigma_{eg}^{+}e^{i\Delta_B t}+b^{\dagger}\sigma_{eg}^{-}e^{-i\Delta_B t})\notag \\
&+&\sqrt{2}g_A(a\sigma_{fe}^{+}e^{i\delta_A t}+a^{\dagger}\sigma_{fe}^{-}e^{-i\delta_A t})\notag \\
&+&\sqrt{2}g_B(b\sigma_{fe}^{+}e^{i\delta_B t}+b^{\dagger}\sigma_{fe}^{-}e^{-i\delta_B t}),
\end{eqnarray}
where the detunings $\Delta_j=\omega_{eg}-\omega_{j}$ and $\delta_j=\omega_{fe}-\omega_{j}=\Delta_j-\alpha$, with the transmon anharmonicity $\alpha=\omega_{eg}-\omega_{fe}>0$  ($j=A,B$). 
The first and second terms describe the off-resonant coupling between cavity $j$
and the $\left| g\right\rangle \leftrightarrow \left| e\right\rangle $
transition of transmon (Fig.~3), 
while the third and fourth terms denote the off-resonant coupling between cavity $%
j $ and the $\left| e\right\rangle \leftrightarrow \left| f\right\rangle $
transition of transmon (Fig.~3), respectively.
For a transmon, a ratio $3\%-5\%$ of the
anharmonicity between the $|g\rangle \leftrightarrow |e\rangle $ transition
frequency and the $|e\rangle \leftrightarrow |f\rangle $ transition
frequency is readily achieved in experiments. In the following,
we choose $|\Delta_j|> \alpha$ to derive the following effective Hamiltonian by using the method \cite{07James}.

Applying the large-detuning conditions $|\Delta_A|\gg g_A$ and $|\Delta_B|\gg g_B$ (i.e., $|\delta_A|\gg \sqrt{2}g_A$ and $|\delta_B|\gg \sqrt{2}g_B$), the Hamiltonian~(\ref{eq102}) changes to \cite{07James}
\begin{eqnarray}  \label{eq103}
H_{e} &=&\left( \frac{2g_A^2}{\delta_A} aa^\dagger+\frac{2g_B^2}{\delta_B}b
b^\dagger\right) \left| f\rangle\langle f\right |\notag \\
&+&\left( \frac{g_A^2}{\Delta_A} aa^\dagger+\frac{g_B^2}{\Delta_B}b
b^\dagger-\frac{2g_A^2}{\delta_A}a^\dagger a-\frac{2g_B^2}{\delta_B}
b^\dagger b\right) \left| e\rangle\langle e\right |\notag \\
 &-&\left( \frac{g_A^2}{\Delta_A}a^\dagger a +
\frac{g_B^2}{\Delta_B}b^\dagger b\right) |g\rangle\langle g|  \notag \\
&+&\frac{g_A g_B}{2}(\frac{1}{\Delta_A}+\frac{1}{\Delta_B})(a
b^\dagger e^{i\Delta_{AB}t}+a^\dagger b e^{-i\Delta_{AB}t})\sigma_z^{eg}\notag \\
&+&g_A g_B(\frac{1}{\delta_A}+\frac{1}{\delta_B})(a
b^\dagger e^{i\delta_{AB}t}+a^\dagger be^{-i\delta_{AB}t})\sigma_z^{fe},
\end{eqnarray}
where $\sigma_z^{eg}=|e\rangle\langle e|-|g\rangle\langle g|$,
 $\sigma_z^{fe}=|f\rangle\langle f|-|e\rangle\langle e|$, $\Delta_{AB}=\Delta_{A}-\Delta_{B}=\omega_B-\omega_A$,
and $\delta_{AB}=\delta_{A}-\delta_{B}=\Delta_{AB}$.
Notice that the first, second, and third lines
of Eq.~(\ref{eq103}) describe Stark shifts of the levels $|f\rangle$, $|e\rangle$, and $|g\rangle$ of the transmon, respectively.
If we eliminate the degrees of freedom of the transmon, the fourth or fifth line
of Eq.~(\ref{eq103}) is exactly the standard Hamiltonian which describes the Jaynes-Cummings interaction between the cavities $A$ and $B$. 
 For simplicity, we set $\Delta_A=\Delta_B=\Delta$ and $g_A=g_B=g$. Therefore, the Hamiltonian~(\ref{eq103}) can be rewritten as
\begin{eqnarray}  \label{eq104}
H_{e}&=&\lambda\left(   a^\dagger a +b^\dagger b+2\right) \left|f\rangle\langle f\right |
+[\Lambda\left( a^\dagger a +b^\dagger b\right)+2\chi] |e\rangle\langle e|  \notag \\
&-&\chi\left( a^\dagger a +b^\dagger b\right) |g\rangle\langle g| 
-\chi\left(a^\dagger b +a
b^\dagger \right)|g\rangle\langle g| \notag \\
&+&\lambda\left(a^\dagger b +a
b^\dagger \right)|f\rangle\langle f|
+\Lambda\left(a^\dagger b +a
b^\dagger \right)|e\rangle\langle e|,
\end{eqnarray}
where the dispersive
shifts $\chi=g^2/\Delta$, $\lambda=2g^2/\delta$, and $\Lambda=\chi-\lambda$, and we have used $[a, a^\dagger]=1$ and $[b, b^\dagger]=1$. Here the dispersive
shifts $\chi$, $\lambda$, and $\Lambda$
can be adjusted by the couplings strength $g$ and the detuning $\Delta$. 
One can obtain the dispersive
shifts $\chi>0$, $\lambda>0$ and $\Lambda<0$ with the detuning $\Delta>0$. On the contrary, if the detuning $\Delta<0$,
we will find that the dispersive
shifts $\chi<0$, $\lambda<0$ and $\Lambda>0$.  

\section{Experimental implementation of
a quantum adder in 3D circuit QED}
\label{sec4}

In this section we will show how to prepare a superposition state for arbitrary input states in 3D circuit QED with the above Hamiltonian with the weak anharmonicity.
Let $|\psi \rangle _{A}=\sum\limits_{n=0}^{\infty }c_{n}|n\rangle _{A}
$ and $|\varphi \rangle _{B}=\sum\limits_{m=0}^{\infty }d_{m}|m\rangle _{B}$ be two arbitrary input pure states
of microwave cavities $A$ and $B$, respectively.  
Here, the Fock states of the cavities are expressed as $|n\rangle _{A}=\frac{(a^{\dagger })^{n}}{\sqrt{n!}}|0\rangle _{A}
$ and $|m\rangle _{B}=\frac{(b^{\dagger })^{m}}{\sqrt{m!}}|0\rangle _{B}$.
In the following, 
we assume that transmon qubit are in arbitrary superposition states $|\phi \rangle
_{T}=\sin\theta |g\rangle +\cos\theta  |f\rangle $, $|\phi \rangle
_{T}=\sin\theta |g\rangle +\cos\theta  |e\rangle $, respectively. 

\subsection{\textbf{ $|e\rangle$ as the auxiliary state} }
We suppose the level $|e\rangle$ of transmon is not occupied, namely, the transmon is initially prepared in state
\begin{eqnarray}  \label{ineq1}
|\phi \rangle
_{T}=\sin\theta |g\rangle +\cos\theta  |f\rangle.
\end{eqnarray}
Therefore, the effective Hamiltonian~(\ref{eq104}) reduces to
\begin{eqnarray}  \label{eq105}
H_{e}&=&\lambda\left(a^\dagger a +b^\dagger b+2\right) \left|f\rangle\langle f\right |
+\lambda\left(a^\dagger b +a
b^\dagger \right)|f\rangle\langle f|  \notag \\
&-&\chi\left( a^\dagger a +b^\dagger b\right) |g\rangle\langle g|
-\chi\left(a^\dagger b +a
b^\dagger \right)|g\rangle\langle g|.
\end{eqnarray}

Under the Hamiltonian~(\ref{eq105}), we obtain from the initial state of the total system $|\phi\rangle_{T} |\psi \rangle _{A}|\varphi \rangle _{B}$
\begin{eqnarray} \label{eq106}
&&e^{-iH_{0}^{g}t}e^{-iH_{I}^{g}t}\sin\theta |g\rangle |\psi \rangle _{A}|\varphi \rangle
_{B}\notag \\
&+&e^{-iH_{0}^{f}t}e^{-iH_{I}^{f}t}\cos\theta|f\rangle |\psi \rangle
_{A}|\varphi \rangle _{B}
\end{eqnarray}
where $H_{0}^{g}=-\chi (a^{\dagger }a+b^{\dagger }b)$,  $H_{I}^{g}=-\chi(a^{\dagger }b+a b^{\dagger })$, $H_{0}^{f}=\lambda\left(a^\dagger a +b^\dagger b+2\right)$,
and $H_{I}^{f}=\lambda\left(a^\dagger b +a
b^\dagger \right)$.

By solving the Heisenberg equations for $H_{I}^{g}$ and $H_{I}^{f}$,
the dynamics of the operators $a^{\dagger }$ and $b^{\dagger }$ can be derived
as
\begin{eqnarray}\label{eq107}
a^{\dagger }(t)&=&\cos (\chi t)a^{\dagger }+i\sin (\chi t)b^\dagger,     \nonumber \\
b^\dagger(t)&=&\cos (\chi
t)b^{\dagger }+i\sin (\chi t)a^\dagger,
\end{eqnarray}
for $|g\rangle$ component and
\begin{eqnarray}\label{ineq3}
a^{\dagger }(t)&=&\cos (\lambda t)a^{\dagger }-i\sin (\lambda t)b^\dagger, \nonumber \\
b^\dagger(t)&=&\cos (\lambda
t)b^{\dagger }-i\sin (\lambda t)a^\dagger,
\end{eqnarray}
for $|f\rangle$ component of the state (\ref{eq106}), respectively. The above Eqs.~(\ref{eq107}) and (\ref{ineq3}) will be used to generate a
superposition state of two cavities.

\subsubsection{\textbf{ $|g\rangle$ as the control state}}
For the evolution time $t=\pm(\frac{\pi}{2}+2k_1\pi)/\chi$,
one has $a^{\dagger }(t)=ib^\dagger$ and $b^\dagger(t)=ia^\dagger$. 
That corresponds to for the level $|g\rangle$, an exchange
of quantum states between two microwave cavities except
for a phase shift $\pi/2$ conditioned on the photon number in cavities $A$ and $B$. When the condition $\lambda
t=\pm(2\pi+2k_2\pi)$ is satisfied, we also have
$a^{\dagger }(t)=a^\dagger$ and $b^\dagger(t)=b^\dagger$
for the level $|f\rangle$. Here, the sign $``+"$ or $``-"$ depends on the detuning $\Delta>0$ or $\Delta<0$, and $k_1$ and $k_2$ are  non-negative integers.  Accordingly, one has the following relationship between the detuning $\Delta$ and anharmonicity $\alpha$
\begin{eqnarray}\label{eq108}
\Delta=\frac{2(k_2+1)}{2k_2-4k_1+1}\alpha
\end{eqnarray}

When the conditions $t=\pm(\frac{\pi}{2}+2k_1\pi)/\chi$ and $\lambda
t=\pm(2\pi+2k_2\pi)$ are satisfied, the Eq.~(\ref{eq106}) changes to (see Eq.~(\ref{eq:er2}) of Appendix \ref{Appendixb})
\begin{eqnarray} \label{eq109}
\sin\theta |g\rangle | \varphi\rangle _{A}|\psi \rangle
_{B}+\cos\theta|f\rangle |\psi \rangle
_{A}|\varphi \rangle _{B}
\end{eqnarray}
where we have used $a^{\dagger }a|n\rangle =n|n\rangle$, $b^{\dagger
}b|m\rangle =m|m\rangle$,
$(i)^{n}=e^{i \frac{\pi}{2} n}$, and $(i)^{m}=e^{i \frac{\pi}{2} m}$. It should be noted here that $ |\psi\rangle$ and $|\varphi\rangle$ have the same Hilbert space while they are arbitrary asymmetric states.

We first apply a  $R_{\pi}^{ef}$ rotation on the transmon that leads to $|f\rangle\rightarrow|e\rangle $. Then 
one performs a  $R_{\pi/2}^{ge}$ 
rotation on the transmon that realizes the conversions $|g\rangle\rightarrow|+\rangle $ and
$|e\rangle\rightarrow|-\rangle $ with $|+\rangle=(|e\rangle +|g\rangle )/\sqrt{2}$ and $|-\rangle=(|e\rangle -|g\rangle )/\sqrt{2}.$
Now we perform a projective measurement onto the state $|+\rangle$ or $|-\rangle$ of 
transmon qubit,  the state~(\ref{eq109}) becomes
\begin{eqnarray}\label{eq110}
\sin\theta|\varphi \rangle
_{A}|\psi \rangle _{B}\pm\cos\theta|\psi \rangle _{A}|\varphi \rangle
_{B}.
\end{eqnarray}
Then we perform another measurement on the cavity $B$ in the referential state $|\chi\rangle_B$ that satisfies $\langle \chi |\psi\rangle_B\neq 0$ and $\langle \chi |\varphi\rangle_B\neq 0$.
Thus, one can obtain the following superposition state of cavity $A$  
\begin{eqnarray}  \label{eq111}
|\Psi\rangle_A= \frac{1}{N}(\gamma|\varphi\rangle\pm\eta|\psi\rangle),
\end{eqnarray}
where $\gamma=\sin\theta\langle\chi\vert\psi\rangle_B$, 
$\eta=\cos\theta\langle\chi\vert\varphi\rangle_B$, and $N=\sqrt{(1/2)[\vert\gamma\vert^2+\vert\eta\vert^2\pm2\textrm{Re}(\gamma \eta^*\langle\psi\vert\varphi\rangle)]}$.
Here, the sign $``+" $or $``-"$ of the output state conditioning on the
measurement $|+\rangle$ or $|-\rangle$ of 
transmon qubit. It can be seen that the performance of the  above superposition state is possible with prior knowledge of
the overlaps of $\langle\chi\vert\varphi\rangle$ and
$\langle\chi\vert\psi\rangle$.

\subsubsection{\textbf{$|f\rangle$ as the control state }}

By waiting for the evolution time $t=\pm(\frac{\pi}{2}+2k_1\pi)/\lambda$,
one has $a^{\dagger }(t)=-ib^\dagger$ and $b^\dagger(t)=-ia^\dagger$. That corresponds to for the level $|f\rangle$, an exchange
of quantum states between two microwave cavities except
for a phase shift $3\pi/2$ conditioned on the photon number in cavities. When the condition $\chi
t=\pm(2\pi+2k_2\pi)$ is satisfied, we also have
$a^{\dagger }(t)=a^\dagger$ and $b^\dagger(t)=b^\dagger$
for the level $|g\rangle$. Here, the sign $``+" $ or $``-"$ depends on the detuning $\Delta>0$ or $\Delta<0$, and $k_1$ and $k_2$ are  non-negative integers.  Accordingly, one has the following relationship between the detuning $\Delta$ and anharmonicity $\alpha$
\begin{eqnarray}\label{eq112}
\Delta=\frac{4k_1+1}{4k_1-8k_2-7}\alpha
\end{eqnarray}

When the conditions $t=\pm(\frac{\pi}{2}+2k_1\pi)/\lambda$ and $\chi
t=\pm(2\pi+2k_2\pi)$ are satisfied, the Eq.~(\ref{eq106})  changes to (see  Eq.~(\ref{eq:er3}) of Appendix \ref{Appendixb})
\begin{eqnarray} \label{eq113}
\sin\theta|g\rangle  |\psi\rangle_A|\varphi\rangle_B 
-\cos\theta|f\rangle |\varphi \rangle
_{A}|\psi \rangle _{B}\end{eqnarray}
where we have used $a^{\dagger }a|n\rangle =n|n\rangle$, $b^{\dagger
}b|m\rangle =m|m\rangle$,
$(-i)^{n}=e^{-i \frac{\pi}{2} n}$, and $(-i)^{m}=e^{-i \frac{\pi}{2} m}$.

We successively apply  $R_{\pi}^{ef}$  and  $R_{\pi/2}^{ge}$ 
rotations to the transmon qubit that result in $|f\rangle\rightarrow|e\rangle\rightarrow|-\rangle$ and $|g\rangle\rightarrow|+\rangle$ with $|+\rangle=(|e\rangle +|g\rangle )/\sqrt{2}$ and $|-\rangle=(|e\rangle -|g\rangle )/\sqrt{2}.$
Now we perform a projective measurement onto the state $|\pm\rangle$ of 
transmon, the state~(\ref{eq113}) becomes
\begin{eqnarray} \label{f1}
\sin\theta|\psi\rangle_A|\varphi\rangle_B 
\mp\cos\theta |\varphi \rangle
_{A}|\psi \rangle _{B}\end{eqnarray}

Then we perform another measurement on the cavity $B$ in the referential state $|\chi\rangle_B$ that satisfies $\langle \chi |\psi\rangle_B\neq 0$ and $\langle \chi |\varphi\rangle_B\neq 0$.
Thus, one obtains the superposition state of cavity $A$  
\begin{eqnarray}  \label{f2}
|\Psi\rangle_A= \frac{1}{N}(\gamma|\psi\rangle\mp\eta|\varphi\rangle),
\end{eqnarray}
where $\gamma=\sin\theta\langle\chi\vert\varphi\rangle_B$, 
$\eta=\cos\theta\langle\chi\vert\psi\rangle_B$, and $N=\sqrt{(1/2)[\vert\gamma\vert^2+\vert\eta\vert^2\mp2\textrm{Re}(\gamma \eta^*\langle\varphi\vert\psi\rangle)]}$.
Here, the sign $``-"$ or $``+"$ of the output state depending on the
measurement $|+\rangle$ or $|-\rangle$ of 
transmon. 

\subsection{\textbf{$|f\rangle$ as the auxiliary state} }

We assume that the level $|f\rangle$ of transmon is not occupied, thus the initial
state of the transmon is
\begin{eqnarray}  \label{ineq2}
|\phi \rangle
_{T}=\sin\theta |g\rangle +\cos\theta  |e\rangle.
\end{eqnarray}
Accordingly,
the effective Hamiltonian Eq.~(\ref{eq104}) reduces to
\begin{eqnarray}  \label{eq114}
H_{e}&=&[\Lambda\left( a^\dagger a +b^\dagger b\right)+2\chi] |e\rangle\langle e|  
+\Lambda\left(a^\dagger b +a
b^\dagger \right)|e\rangle\langle e|\notag \\
&-&\chi\left( a^\dagger a +b^\dagger b\right) |g\rangle\langle g| 
-\chi\left(a^\dagger b +a
b^\dagger \right)|g\rangle\langle g|.
\end{eqnarray}
According to the Hamiltonian~(\ref{eq114}), the initial state
$|\phi\rangle_{T} |\psi \rangle _{A}|\varphi \rangle _{B}$ of the total system at time $t$ becomes
\begin{eqnarray} \label{eq115}
&&e^{-iH_{0}^{g}t}e^{-iH_{I}^{g}t}\sin\theta |g\rangle |\psi \rangle _{A}|\varphi \rangle
_{B}\notag \\
&+&e^{-iH_{0}^{e}t}e^{-iH_{I}^{e}t}\cos\theta|e\rangle |\psi \rangle
_{A}|\varphi \rangle _{B}
\end{eqnarray}
where $H_{0}^{g}=-\chi (a^{\dagger }a+b^{\dagger }b)$,  $H_{I}^{g}=-\chi(a^{\dagger }b+a b^{\dagger })$, $H_{0}^{e}=\Lambda\left( a^\dagger a +b^\dagger b\right)+2\chi $,
and $H_{I}^{e}=\Lambda\left(a^\dagger b +a
b^\dagger \right)$. 

By solving the Heisenberg equations for $H_{I}^{g}$ and $H_{I}^{e}$,
the dynamics of the operators $a^{\dagger }$ and $b^{\dagger }$ can be derived
as
\begin{eqnarray}\label{eq116}
a^{\dagger }(t)&=&\cos (\chi t)a^{\dagger }+i\sin (\chi t)b^\dagger,     \nonumber \\
b^\dagger(t)&=&\cos (\chi
t)b^{\dagger }+i\sin (\chi t)a^\dagger, 
\end{eqnarray}
for $|g\rangle$ component, and 
\begin{eqnarray}\label{ineq4}
a^{\dagger }(t)&=&\cos (\Lambda t)a^{\dagger }-i\sin (\Lambda t)b^\dagger, \nonumber \\
b^\dagger(t)&=&\cos (\Lambda
t)b^{\dagger }-i\sin (\Lambda t)a^\dagger,
\end{eqnarray}
for $|e\rangle$ component. We then discuss how to use the Eqs.~(\ref{eq116}) and (\ref{ineq4}) to realize a superposition state of two cavities.

\subsubsection{\textbf{$|g\rangle$ as the control state} }
For the evolution time $t=\pm(\frac{\pi}{2}+2k_1\pi)/\chi$,
one has $a^{\dagger }(t)=ib^\dagger$ and $b^\dagger(t)=ia^\dagger$. That corresponds to for the level $|g\rangle$, an exchange
of quantum states between two microwave cavities except
for a phase shift $\pi/2$ conditioned on the photon number in cavities. When the condition $\Lambda
t=\mp(2\pi+2k_2\pi)$ is satisfied, we also have
$a^{\dagger }(t)=a^\dagger$ and $b^\dagger(t)=b^\dagger$
for the level $|e\rangle$. Accordingly, we can obtain the following relationship between the $\Delta$ and anharmonicity $\alpha$
\begin{eqnarray}\label{eq117}
\Delta=\frac{4k_1+4k_2+5}{4k_2-4k_1+3}\alpha
\end{eqnarray}
When the conditions $t=(\frac{\pi}{2}+2k_1\pi)/|\chi|$ and $|\Lambda|
t=2\pi+2k_2\pi$ are satisfied, the Eq.~(\ref{eq115}) changes to (see  Eq.~(\ref{eq:er4}) of Appendix  \ref{Appendixc})
\begin{eqnarray} \label{eq118}
\sin\theta |g\rangle | \varphi\rangle _{A}|\psi \rangle
_{B}-\cos\theta|e\rangle |\psi \rangle
_{A}|\varphi \rangle _{B}
\end{eqnarray}
where we have used $a^{\dagger }a|n\rangle =n|n\rangle$, $b^{\dagger
}b|m\rangle =m|m\rangle$,
$(i)^{n}=e^{i \frac{\pi}{2} n}$, and $(i)^{m}=e^{i \frac{\pi}{2} m}$. Here, the sign $\pm$ of Eq.~(3) depending on the detuning $\Delta$. 
It should be noted here that $ |\psi\rangle|$ and $|\varphi\rangle$ have the same Hilbert space while they are arbitrary asymmetric states.

We perform a  $R_{\pi/2}^{ge}$ 
rotation on the transmon qubit that realizes the conversions $|g\rangle\rightarrow|+\rangle $ and
$|e\rangle\rightarrow|-\rangle $ with $|+\rangle=(|e\rangle +|g\rangle )/\sqrt{2}$ and $|-\rangle=(|e\rangle -|g\rangle )/\sqrt{2}.$
Now we perform a projective measurement onto the state $|+\rangle$ or $|-\rangle$ of 
transmon qubit, the state~(\ref{eq118}) becomes
\begin{eqnarray}\label{eq130}
\sin\theta |g\rangle | \varphi\rangle _{A}|\psi \rangle
_{B}\mp\cos\theta|e\rangle |\psi \rangle
_{A}|\varphi \rangle _{B}.
\end{eqnarray}
Then we perform another measurement on the cavity $B$ in the referential state $|\chi\rangle_B$ that satisfies $\langle \chi |\psi\rangle_B\neq 0$ and $\langle \chi |\varphi\rangle_B\neq 0$.
Thus, one can obtain the following superposition state of cavity $A$  

\begin{eqnarray}  \label{eq131}
|\Psi\rangle_A= \frac{1}{N}(\gamma|\varphi\rangle\mp\eta|\psi\rangle),
\end{eqnarray}
where $\gamma=\sin\theta\langle\chi\vert\psi\rangle_B$, 
$\eta=\cos\theta\langle\chi\vert\varphi\rangle_B$, and $N=\sqrt{(1/2)[\vert\gamma\vert^2+\vert\eta\vert^2\mp2\textrm{Re}(\gamma \eta^*\langle\psi\vert\varphi\rangle)]}$.
Here, the sign $``+" $or $``-"$ of the output state conditioning on the
measurement $|+\rangle$ or $|-\rangle$ of 
transmon qubit. It can be seen that the performance of the  above superposition state is possible with prior knowledge of
the overlaps of $\langle\chi\vert\varphi\rangle$ and
$\langle\chi\vert\psi\rangle$.

\subsubsection{\textbf{$|e\rangle$ as the control state} }
For the evolution time $t=\mp(\frac{\pi}{2}+2k_1\pi)/\Lambda$,
one has $a^{\dagger }(t)=-ib^\dagger$ and $b^\dagger(t)=-ia^\dagger$. 
That corresponds to for the level $|e\rangle$, an exchange
of quantum states between two microwave cavities except
for a phase shift $3\pi/2$ conditioned on the photon number in cavities. Here the sign $``-"$ of the expression $t$
corresponds to $\Delta>0$ while the sign $``+"$ corresponds to $\Delta<0$. When the condition $\chi
t=\pm(2\pi+2k_2\pi)$ is satisfied, we also have
$a^{\dagger }(t)=a^\dagger$ and $b^\dagger(t)=b^\dagger$
for the level $|g\rangle$. Accordingly, one has the following relationship between the detuning $\Delta$ and anharmonicity $\alpha$
\begin{eqnarray}\label{eq132}
\Delta=\frac{4k_1+4k_2+5}{4k_2-4k_1\mp3}\alpha
\end{eqnarray}
When the conditions $t=(\frac{\pi}{2}+2k_1\pi)/|\Lambda|$ and 
$|\chi|t=2\pi+2k_2\pi$ are satisfied, the Eq.~(\ref{eq115}) changes to  (see  Eq.~(\ref{eq:er5}) of Appendix  \ref{Appendixc})
\begin{eqnarray} \label{eq133}
\sin\theta |g\rangle |\psi \rangle
_{A}|\varphi \rangle _{B}+\cos\theta|e\rangle | \varphi\rangle _{A}|\psi \rangle
_{B}
\end{eqnarray}
where we have used $a^{\dagger }a|n\rangle =n|n\rangle$, $b^{\dagger
}b|m\rangle =m|m\rangle$,
$(-i)^{n}=e^{-i \frac{\pi}{2} n}$, and $(-i)^{m}=e^{-i \frac{\pi}{2} m}$. 

We perform a  $R_{\pi/2}^{ge}$ 
rotation on the transmon qubit that realizes the conversions $|g\rangle\rightarrow|+\rangle $ and
$|e\rangle\rightarrow|-\rangle $ with $|+\rangle=(|e\rangle +|g\rangle )/\sqrt{2}$ and $|-\rangle=(|e\rangle -|g\rangle )/\sqrt{2}.$
Now we perform a projective measurement onto the state $|\pm\rangle$ of 
transmon, the state~(\ref{eq133}) becomes
\begin{eqnarray}\label{eq134}
\cos\theta|\psi \rangle _{A}|\varphi \rangle
_{B}}\pm\sin\theta|\varphi \rangle
_{A}|\psi \rangle _{B.
\end{eqnarray}
Then we perform another measurement on the cavity $B$ in the referential state $|\chi\rangle_B$ that satisfies $\langle \chi |\psi\rangle_B\neq 0$ and $\langle \chi |\varphi\rangle_B\neq 0$.
Thus, one can obtain the following superposition state of cavity $A$  
\begin{eqnarray}  \label{eq136}
|\Psi\rangle_A= \frac{1}{N}(\gamma|\psi\rangle\pm\eta|\varphi\rangle),
\end{eqnarray}
where $\gamma=\sin\theta\langle\chi\vert\varphi\rangle_B$, 
$\eta=\cos\theta\langle\chi\vert\psi\rangle_B$, and $N=\sqrt{(1/2)[\vert\gamma\vert^2+\vert\eta\vert^2\pm2\textrm{Re}(\gamma \eta^*\langle\varphi\vert\psi\rangle)]}$.
Here, the sign $``\pm"$ of the output state depending on the
measurement $|\pm\rangle$ of 
transmon.

\section{Possible experimental implementation}
\label{sec5}

Recent experimental results for the 3D circuit system
demonstrate the great promise of
quantum computation and QIP.
For an experimental implementation, our setup of two
superconducting 3D cavities coupled to a transmon has been
demonstrated recently by \cite{16Wangc}. 

Taking into account the effect of transmon weak anharmonicity,
the dissipation and the dephasing, the dynamics
of the lossy system is governed by the Markovian
master equation
\begin{eqnarray}  \label{eq17}
\frac{d\rho }{dt} &=&-i[H_I,\rho ] +\kappa _{A} \mathcal{D}
[a]+\kappa _{B} \mathcal{D}[b]  \notag \\
&+&\gamma _{eg}\mathcal{D}[ \sigma _{eg}^{-}] +\gamma _{fe}\mathcal{D}[
\sigma _{fe}^{-}]+\gamma _{fg}\mathcal{D}[ \sigma _{fg}^{-}]  \notag \\
&+&\gamma _{\varphi e}\mathcal{D}[ \sigma _{ee}]+\gamma _{\varphi f}\mathcal{D}[ \sigma _{ff}]  ,
\end{eqnarray}
where ${\rho }$ is the density matrix of the whole system, ${H_I}$ is given by Eq.~(\ref{eq102}), $\sigma _{ee}=\left\vert e\right\rangle\left\langle e\right\vert, \sigma _{ff}=\left\vert f\right\rangle\left\langle f\right\vert$, and $%
\mathcal{D}\left[ \mathcal{O} \right] =(2\mathcal{O} \rho \mathcal{O} ^{+}-\mathcal{O}
^{+}\mathcal{O} \rho -\rho \mathcal{O} ^{+}\mathcal{O})/2$ is the dissipator. Here, $\kappa_{A}~(\kappa_{B})$ is the
decay rate of cavity $A$ ($B$). In addition, $\gamma _{eg}$,
 $\gamma _{fe}$, and $\gamma _{fg}$ are the energy relaxation rates from state $\left\vert e\right\rangle $ to $\left\vert g\right\rangle$, $\left\vert f\right\rangle $ to $\left\vert e\right\rangle$, and $\left\vert f\right\rangle $ to $\left\vert g\right\rangle$ of
transmon qubit, respectively. $\gamma _{\varphi e}$ ($\gamma _{\varphi f}$)  is the dephasing rate of the
level $\left\vert e\right\rangle $ ($\left\vert f\right\rangle $) of transmon qubit.

The generation efficiency can be evaluated by fidelity $\mathcal{F}=\sqrt{\left\langle \psi _{\mathrm{id}}\right\vert \rho
\left\vert \psi _{\mathrm{id}}\right\rangle}$, where $\left\vert \psi _{\mathrm{id}}\right\rangle $ is the ideal
target state. 
It is obvious that how to realize the controlled-SWAP gate is the key of our proposal. So we mainly consider
the impurity introduced in this process. 
In fact, the initial state preparation and measurement of transmon and cavities can be relatively accurately performed \cite{16Wangc,16Rosenblum,ini1,ini2},  thus it is also reasonable for us not  to consider the initial state preparation, the rotations and measurement impurites on the fidelity. Accordingly,  the ideal target state can just use the states like Eq.~(\ref{eq109}) .
In addition, the input state of the transmon-cavity system is $ |\phi \rangle
_{T} |\psi \rangle _{A}|\varphi \rangle _{B}$.
The initial state of cavities $|\psi \rangle _{A}$ and $|\varphi\rangle _{B}$ can be arbitrary
states such as discrete-variable states or continuous-variable
states. In the following, we choose the
cavities
are initially in the coherent states, i.e.,
$|\psi \rangle _{A}=|\alpha \rangle _{A}$ and $|\varphi \rangle _{B}=|-\beta \rangle _{B}$ with $\alpha=\beta=0.1$. 

We numerically simulate the fidelity of the operation by solving the master equation (\ref{eq17}). Since the transmon qubit relaxation time $T_1=75~\mu s$ and
the transmon qubit dephasing time $T_2=45~\mu s$ have been achieved in similar 3D circuit system \cite{16Wangc}, 
we set 
$\gamma^{-1}_{\varphi e}=15~\mu s,\gamma^{-1}_{\varphi f}=10~\mu s$,
$\gamma^{-1}_{eg}=50~\mu s, \gamma^{-1}_{fe}=25~\mu s, \gamma^{-1}_{fg}=100~\mu s,$ In addition, we set $\kappa^{-1}_{A}=1.5k~\mu s,$ $\kappa^{-1}_{B}=1.0k~\mu s.$
According to Ref.~\cite{16Wangc}, we choose  the transmon anharmonicity $\alpha=\omega_{eg}-\omega_{fe}=115$ MHz in our numerical simulation. 
In the following, we 
choose the dispersive shift $\chi=1$ MHz that because this value of $\chi$ is readily available in experiments \cite{16Wangc}.
\subsection{\textbf{$|e\rangle$ as the auxiliary state} }

In Fig.~4-6 we consider the level $|e \rangle$ of transmon is not occupied. Thus the input state of the transmon-cavity system is $(\sin\theta |g\rangle +\cos\theta  |f\rangle)|\alpha \rangle _{A}|-\beta \rangle _{B}$.  The ideal target state is chosen by Eq.~(\ref{eq109}) or Eq.~(\ref{eq113}) 
for the case of (\romannumeral1) the state $|g\rangle$ acts as a control state or
(\romannumeral2) the state $|f\rangle$ acts as a control state.
The corresponding parameters used in this subsection are: (\romannumeral1) $k_1=1,$ $k_2=2,$ i.e., $\Delta=6\alpha=0.69$ GHz, and (\romannumeral2) $k_1=2,$ $k_2=0,$ i.e., $\Delta=9\alpha=1.035$ GHz, respectively.
\begin{figure}[tbp]
\begin{center}
\subfigure[]{\includegraphics[width=3in]{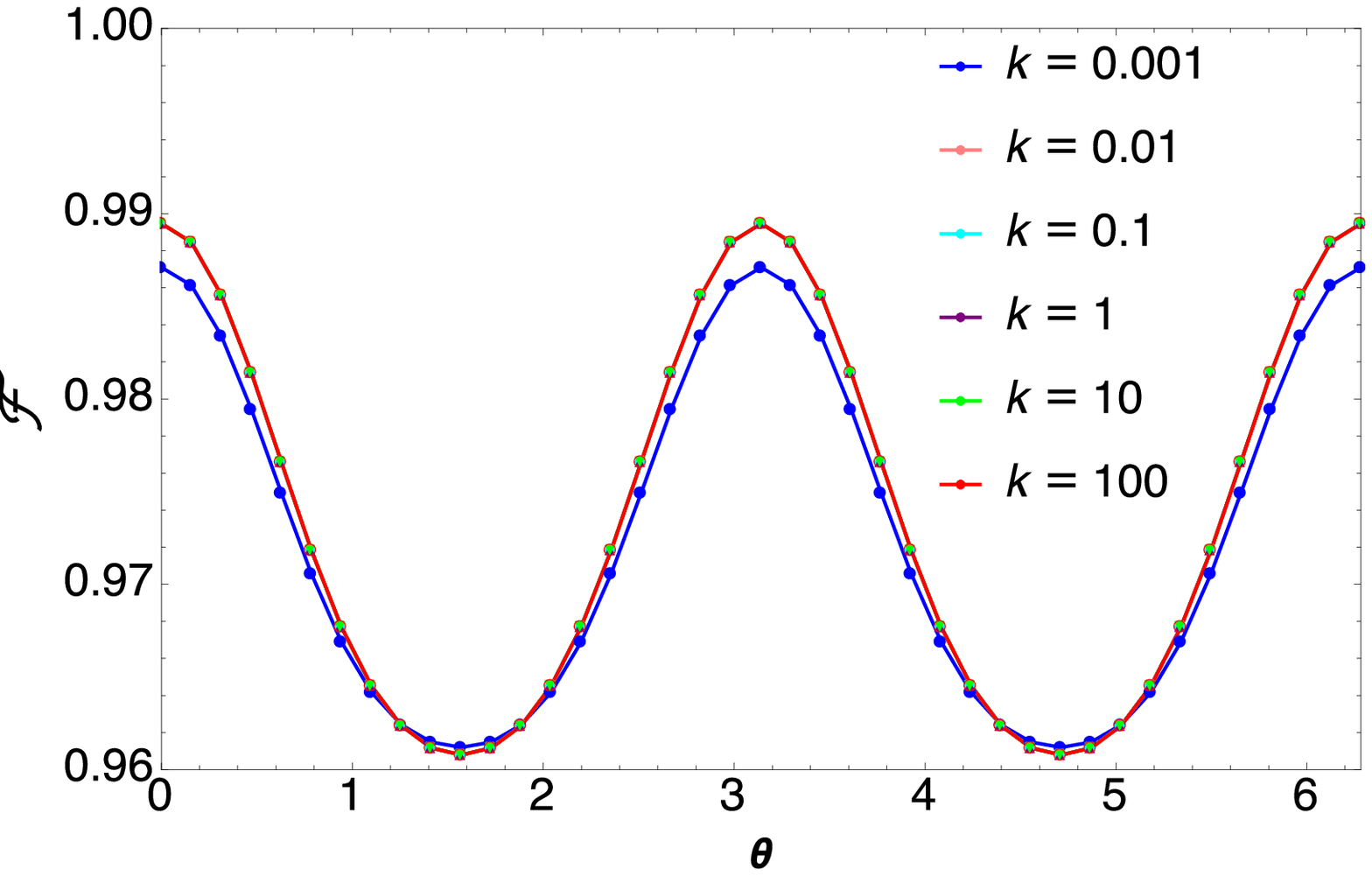}}
\subfigure[]{\includegraphics[width=3in]{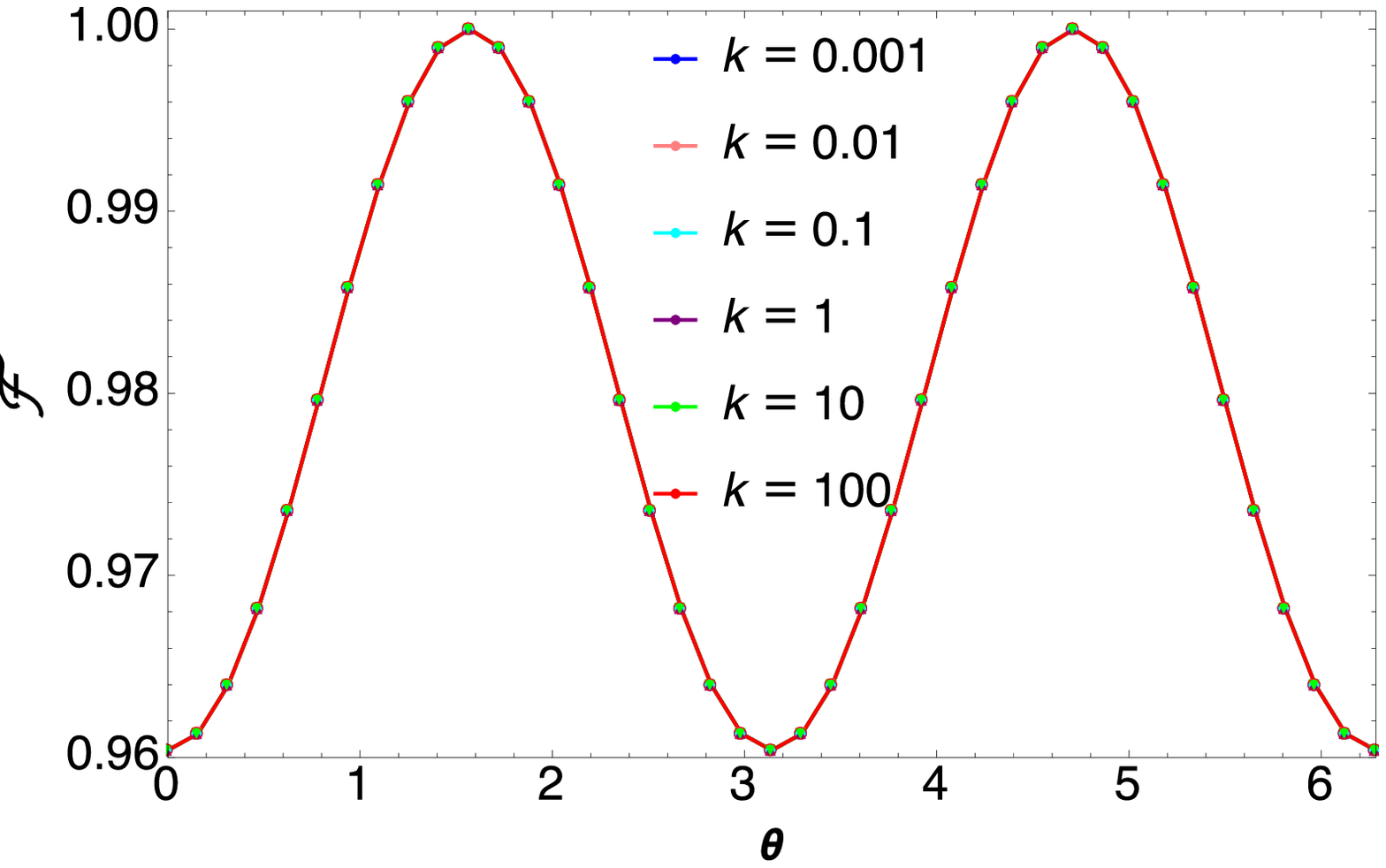}}
\end{center}
\caption{(color online) Fidelity $\mathcal{F}$ versus $\theta$ 
for the case of  (a) the state $|g\rangle$ acts as a control state and
(b) the state $|f\rangle$ acts as a control state for
$k=0.001$,$~0.01$,$~0.1$,$~1$,$~10$,$~100$.
The parameters used in the numerical simulation are referred in the text.}
\label{fig:}
\end{figure}
Figure~4(a) (b) show the fidelity $\mathcal{F}$ versus $\theta$ ($\theta\in[0,2\pi]$) for the case of  (\romannumeral1) the state $|g\rangle$ acts as a control state and
(\romannumeral2) the state $|f\rangle$ acts as a control state, respectively. Fig.~4 (a) or  (b) shows that for $k=0.001,~0.01,~0.1,~1,~10,~100$, the operational fidelity  can be greater than $96.07\%$ or $96.04\%$. 
Moreover, we calculate the average fidelities are approximately (\romannumeral1) $97.31\%$, $97.43\%$, $97.43\%$, $97.43\%$, $97.43\%$, and $97.43\%$, (\romannumeral2) $97.90\%$, $97.90\%$, $97.90\%$, $97.90\%$, $97.90\%$, and $97.90\%$ for $k=0.001,~0.01,~0.1,~1,~10,~100$, respectively.
 As in Fig. 4, the effect of the cavity decay on the fidelity is almost unaffected with the current parameter values.
Thus the decoherence caused by the cavity decay can be greatly suppressed.
In Fig. 5 and Fig. 6, we choose $k=10$ which corresponding to  $\kappa^{-1}_{A}=15~\mu s,$ $\kappa^{-1}_{B}=10~\mu s.$  Recently, the cavity relaxation time $T_1=3.3~m s$  has been reported in 3D circuit system \cite{16Wangc}.

\begin{figure}[tbp]
\begin{center}
\subfigure[]{\includegraphics[width=3in]{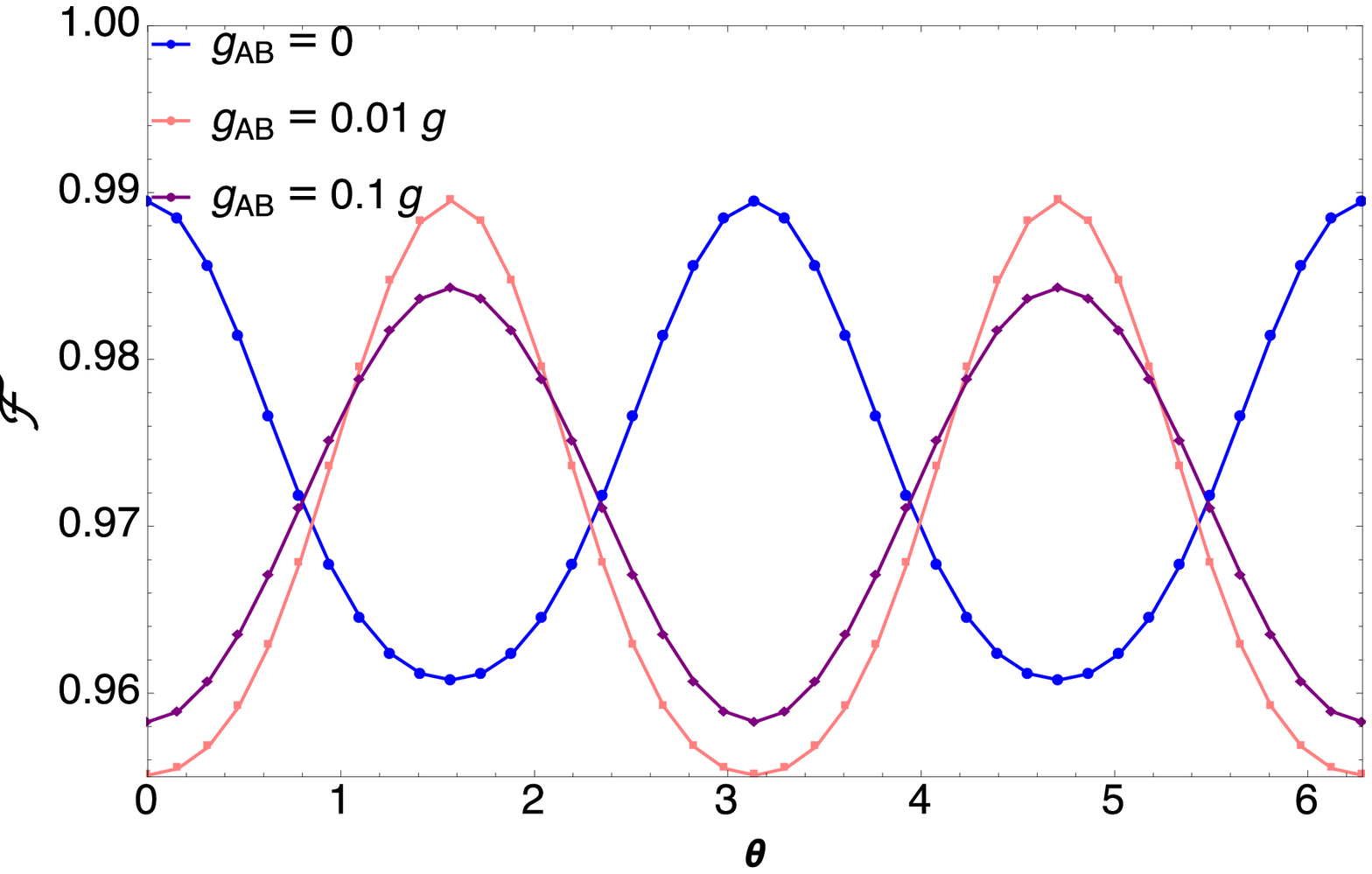}}
\subfigure[]{\includegraphics[width=3in]{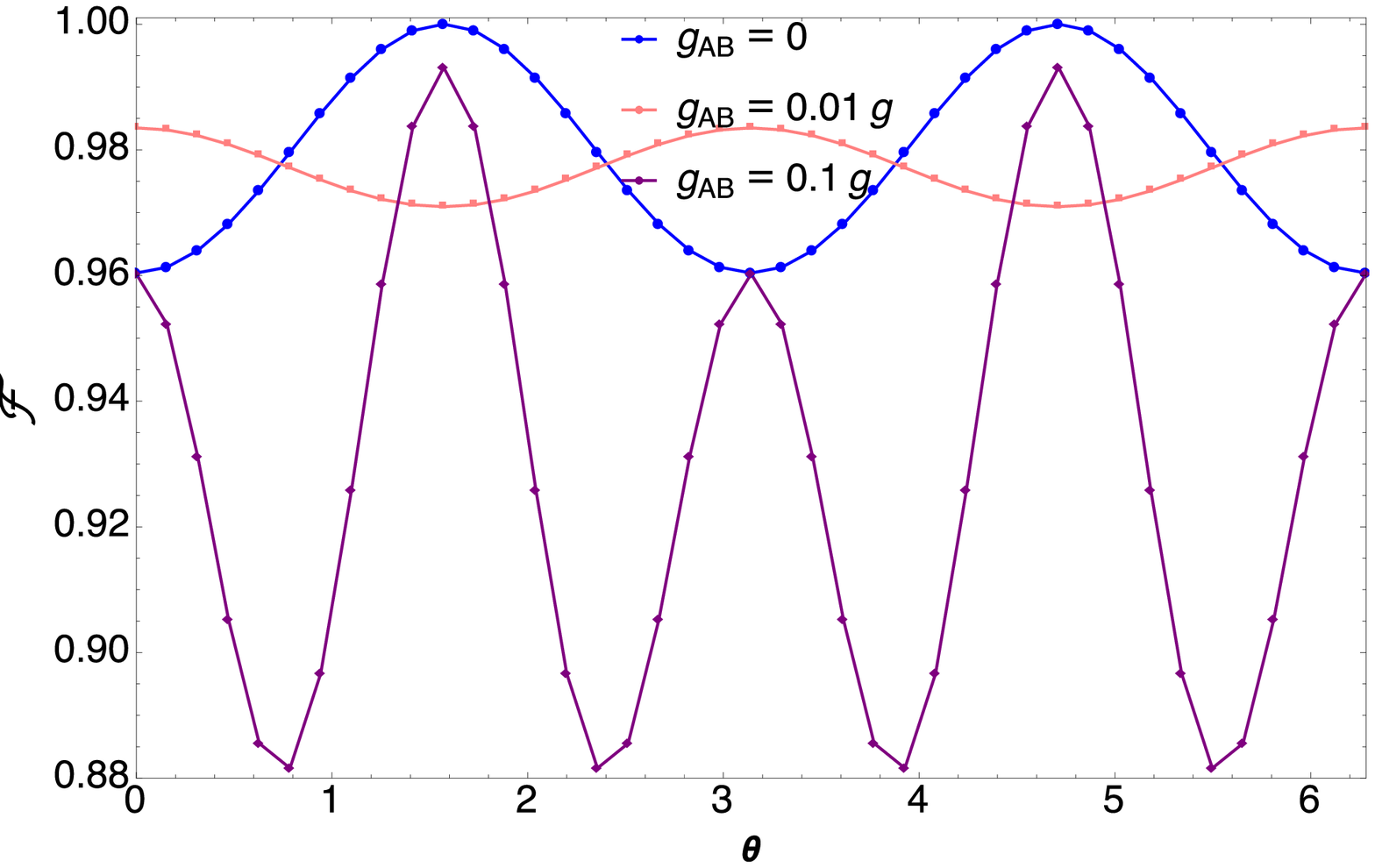}}
\end{center}
\caption{(color online) Fidelity $\mathcal{F}$ versus $\theta$  and by taking the unwanted inter-cavity crosstalk into account for $g_{AB}=0,~0.01g,~0.1g$. Here, the $g_{AB}$  represents the inter-cavity crosstalk coupling strength.}
\label{fig:}
\end{figure}
To investigate the effect of the undesired inter-cavity crosstalk on the fidelity, we numerically calculate the operation fidelity with the crosstalk between two cavities in Fig.~5 (a) (b).
The effect of the inter-cavity crosstalk can be taken into 
account by adding a Hamiltonian of the form $H_{AB}=g_{AB}(a^{\dagger }b+ab^{\dagger })$ in $H_I$,
where $g_{AB}$ is the inter-cavity crosstalk coupling strength. Figures 5 (a) and (b) display fidelity $\mathcal{F}$ versus $\theta$, which are plotted by choosing $g_{AB}=0,~0.01g,~0.1g$
for
the case of  (\romannumeral1) the state $|g\rangle$ acts as a control state and
(\romannumeral2) the state $|f\rangle$ acts as a control state, respectively.
From the Fig.~5 (a)  one
can see that the crosstalk effect is very small or negligible  for $g_{AB}=0.01g,~0.1g$.
One calculates the average fidelities are approximately $97.43\%$, $97.06\%$, and $96.93\%$
for $g_{AB}=0,~0.01g,~0.1g$.
Figure 5(b) shows that for $g_{AB}=0.01g$, the fidelity is almost unaffected by the  crosstalk error while for $g_{AB}=0.1g$, the fidelity has a small decrease.
The corresponding average fidelities $\sim$ $97.90\%$, $97.75\%$, and $93.12\%$  are calculated for  $g_{AB}=0.01g,~0,~0.1g$. In Fig.~6 (a) and 6 (b), we choose  the inter-cavity crosstalk coupling strengths
$g_{AB}=0.1g$ and $g_{AB}=0.01g$, respectively.
These the coupling strength conditions are easily satisfied by the present circuit QED technology \cite{16sr2}.

\begin{figure}[tbp]
\begin{center}
\subfigure[]{\includegraphics[width=3.5in]{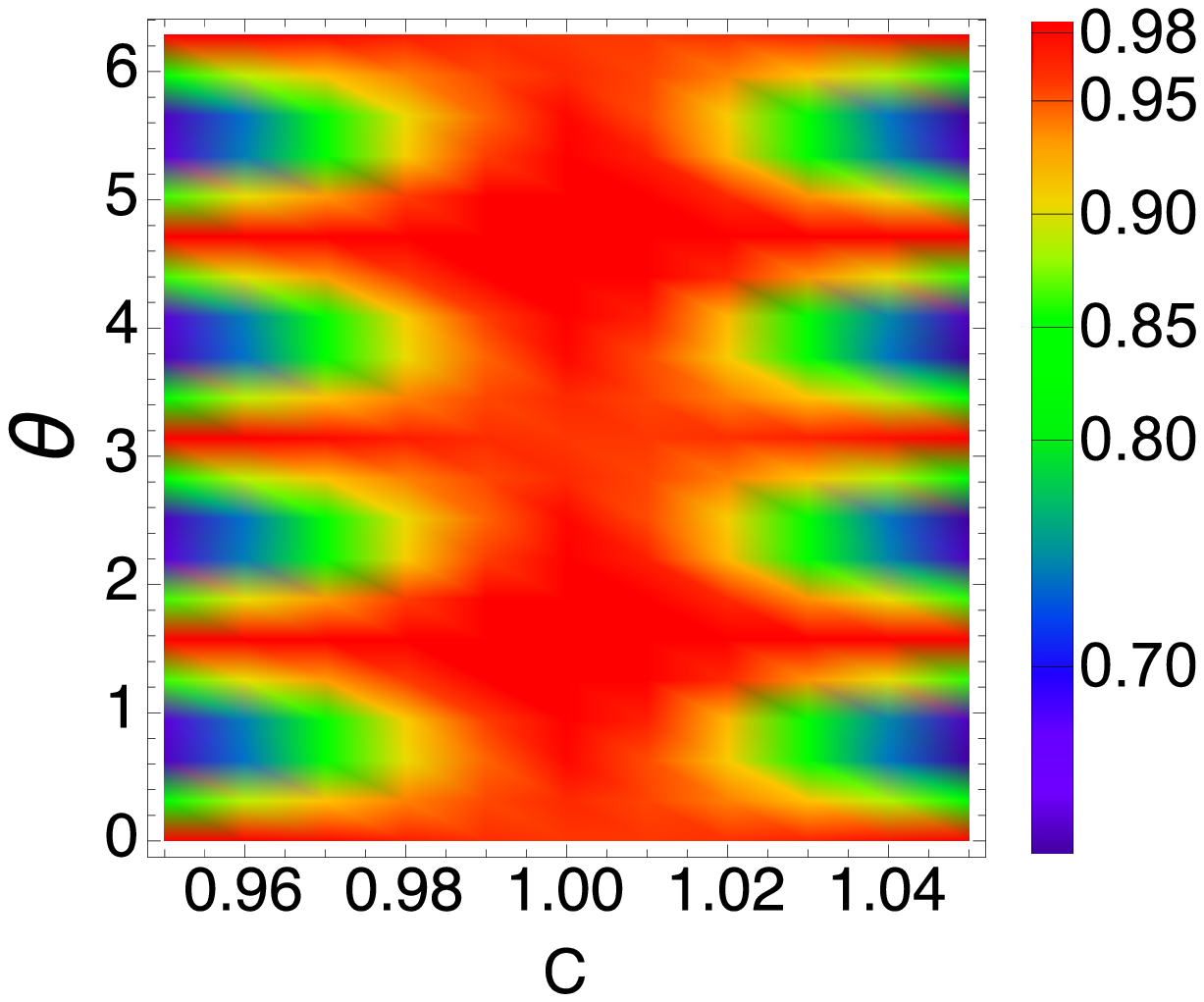}}
\subfigure[]{\includegraphics[width=3.5in]{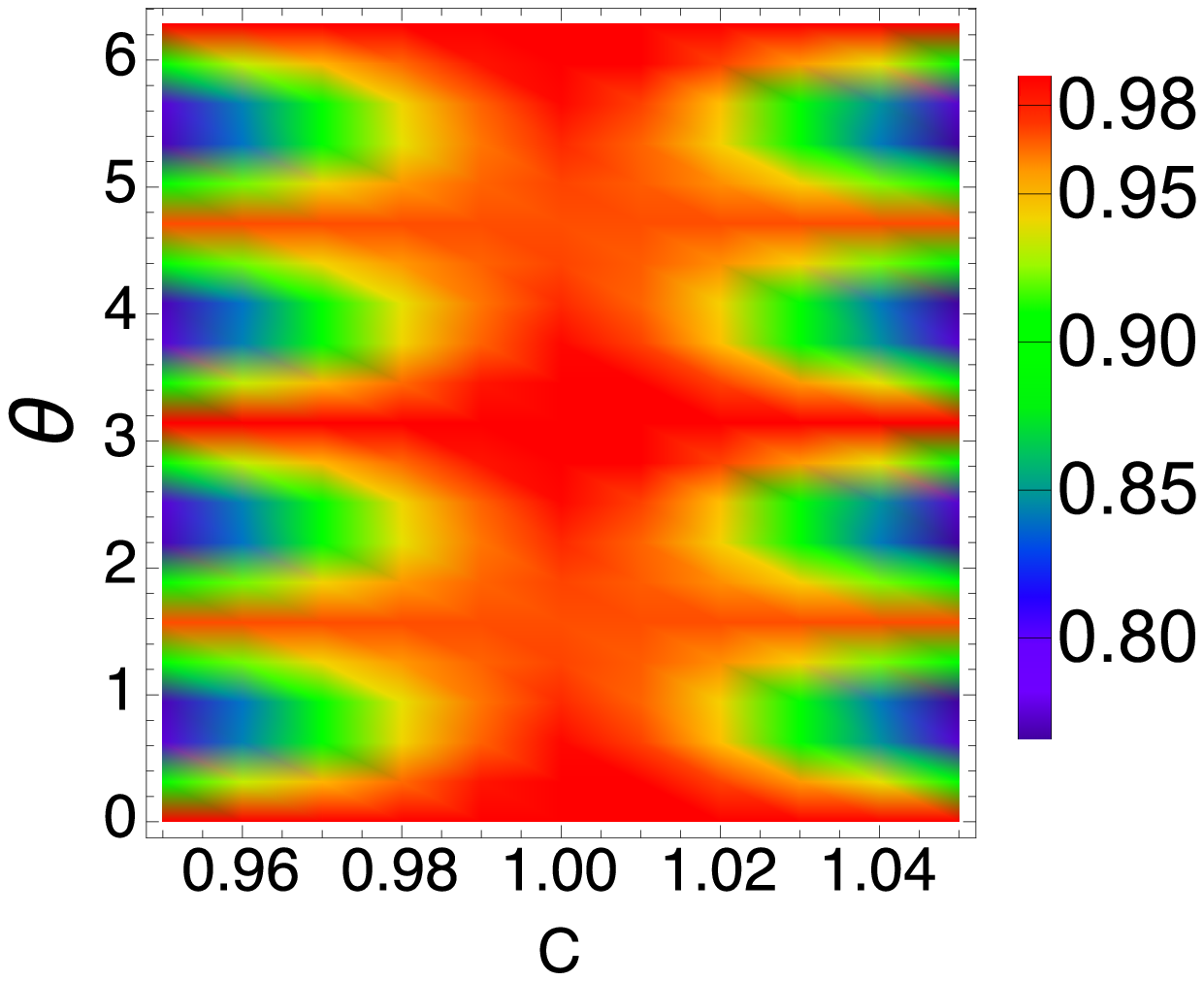}}
\end{center}
\caption{(color online) Fidelity $\mathcal{F}$ versus $c$ and $\theta$ by taking into account the inhomogeneity in transmon-cavity interaction. Here $g_A=g$ and $g_B=cg$ with $c\in[0.95,1.05]$.}
\label{fig:}
\end{figure}

Figure 6 (a) and 6 (b) show output fidelities from the master equation
simulation, which take into account the inhomogeneity in transmon-cavity interaction for 
the case of  (\romannumeral1) the state $|g\rangle$ acts as a control state and
(\romannumeral2) the state $|f\rangle$ acts as a control state, respectively. Figure~6 displays the fidelity versus $c$ and $\theta$, where we set
$g_A=g$ and $g_B=cg$ with $c\in[0.95,1.05]$.
We can see from Fig. 6 (a) that with the change of the $c$,
the fidelity is almost unaffected in the regions $0.99\leq c <1$ and $1< c\leq 1.01$.
Figure 6 (b) shows that the effect of the inhomogeneous coupling on the fidelity is very small for  $0.99\leq c <1$ and $1< c\leq 1.01$.
Figure 6 (a) and 6 (b) mean that the coupling strengths $g_A$ and $g_B$ are not necessarily strictly equal in our proposal.

\subsection{\textbf{$|f\rangle$ as the auxiliary state} }
In this subsection we consider the level $|f \rangle$ of transmon is not occupied, i.e., the input state of the transmon-cavity system is $(\sin\theta |g\rangle +\cos\theta  |e\rangle)|\alpha \rangle _{A}|-\beta \rangle _{B}$.  The ideal target state is given by Eq.~(\ref{eq118}) or Eq.~(\ref{eq133})
for the case of (\romannumeral1) the state $|g\rangle$ acts as a control state or
(\romannumeral2) the state $|e\rangle$ acts as a control state.
The corresponding parameters used in this subsection are: (\romannumeral1) $k_1=1,$ $k_2=0,$ i.e., $\Delta=-9\alpha=-1.035$ GHz, and (\romannumeral2) $k_1=1,$ $k_2=0,$ i.e., $\Delta=-9\alpha=-1.035$ GHz, respectively.
\begin{figure}[tbp]
\begin{center}
\subfigure[]{\includegraphics[width=3in]{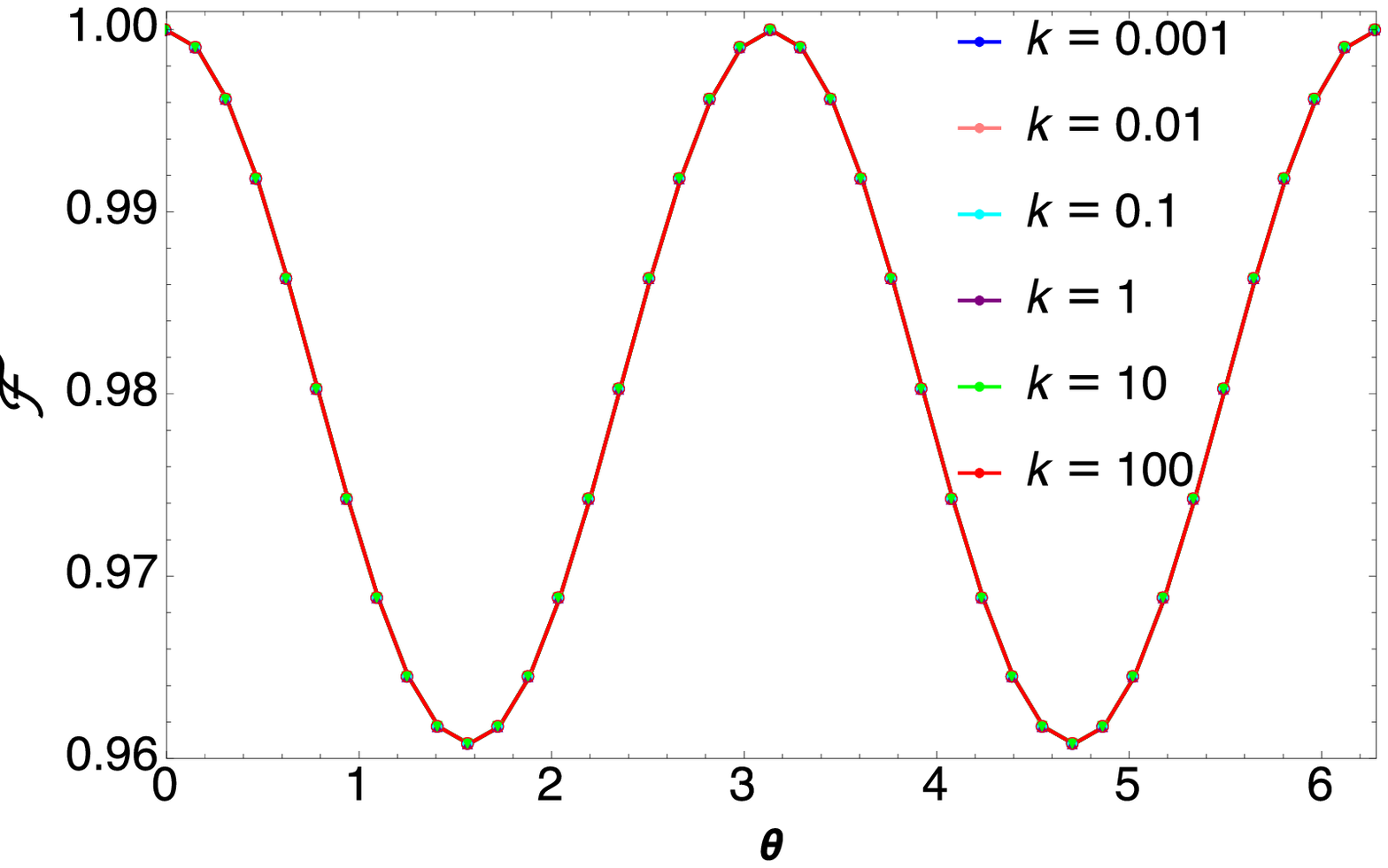}}
\subfigure[]{\includegraphics[width=3in]{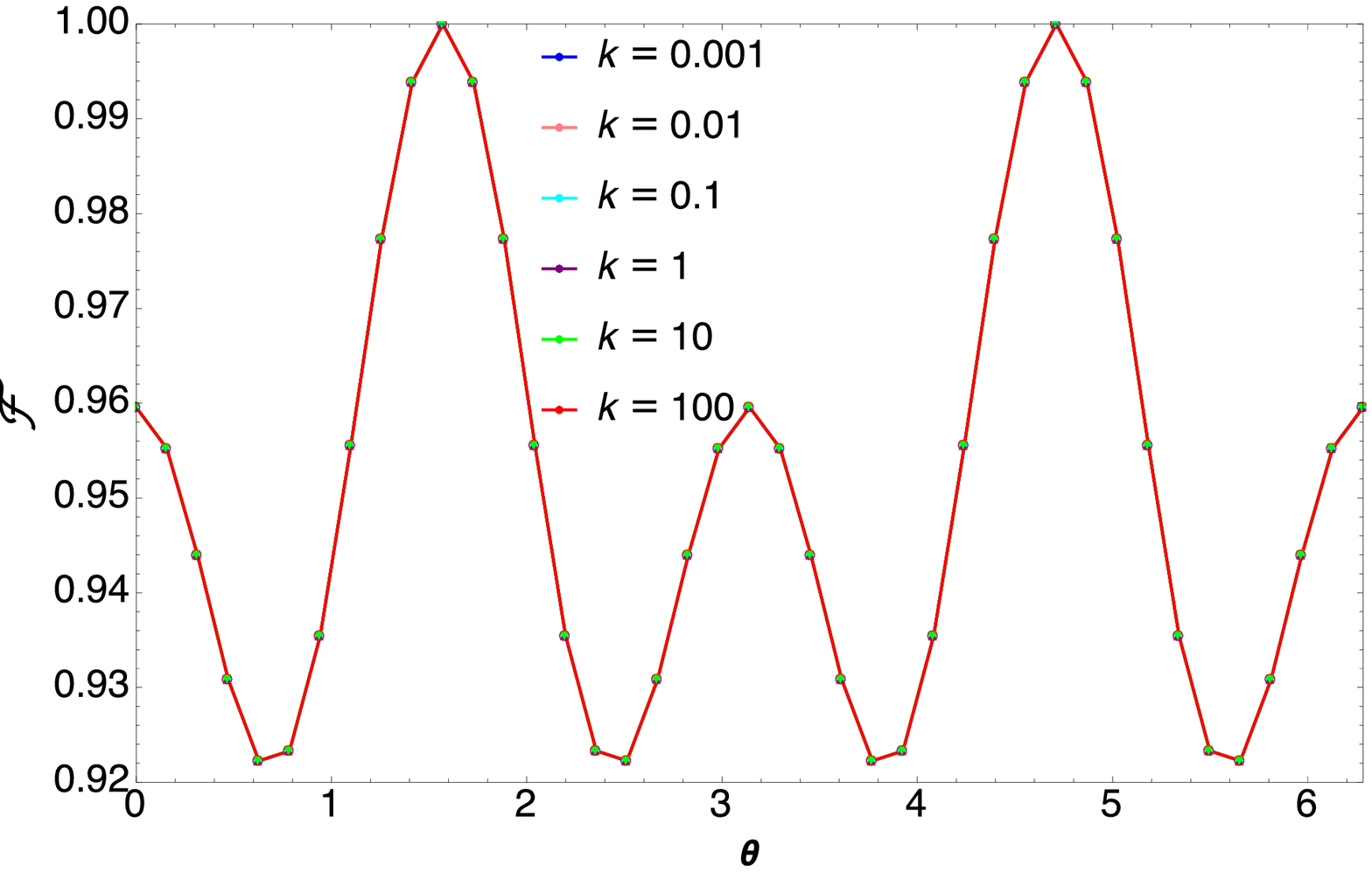}}
\end{center}
\caption{(color online) Fidelity $\mathcal{F}$ versus $\theta$ 
for the case of  (a) the state $|g\rangle$ acts as a control state and
(b) the state $|e\rangle$ acts as a control state for
$k=0.001$,$~0.01$,$~0.1$,$~1$,$~10$,$~100$.
The parameters used in the numerical simulation are referred in the text.}
\label{fig:}
\end{figure}

Figure~7 (a) and (b) display the fidelity $\mathcal{F}$ versus $\theta$ with $\theta\in[0,2\pi]$ for the case of  (\romannumeral1) the state $|g\rangle$ acts as a control state and
(\romannumeral2) the state $|e\rangle$ acts as a control state, respectively. Figure 7 (a) or  (b) shows that for $k=0.001,~0.01,~0.1,~1,~10,~100$, the operational fidelity  can be greater than $96.06\%$ or $92.22\%$. 
In addition, we calculate the average fidelities are approximately (\romannumeral1) 
$98.13\%$, $98.13\%$, $98.13\%$, $98.13\%$, $98.13\%$, and $98.13\%$, 
(\romannumeral2) $95.21\%$, $95.21\%$, $95.21\%$, $95.21\%$, $95.21\%$, and $95.21\%$ for $k=0.001,~0.01,~0.1,~1,~10,~100$, respectively.
 Figure 7 shows the effect of the cavity decay on the fidelity is almost unaffected with the current parameter values.
Thus the decoherence caused by the cavity decay can be greatly suppressed.
In the following, we choose $k=10$ which corresponding to  $\kappa^{-1}_{A}=15~\mu s,$ $\kappa^{-1}_{B}=10~\mu s.$ 

\begin{figure}[tbp]
\begin{center}
\subfigure[]{\includegraphics[width=3in]{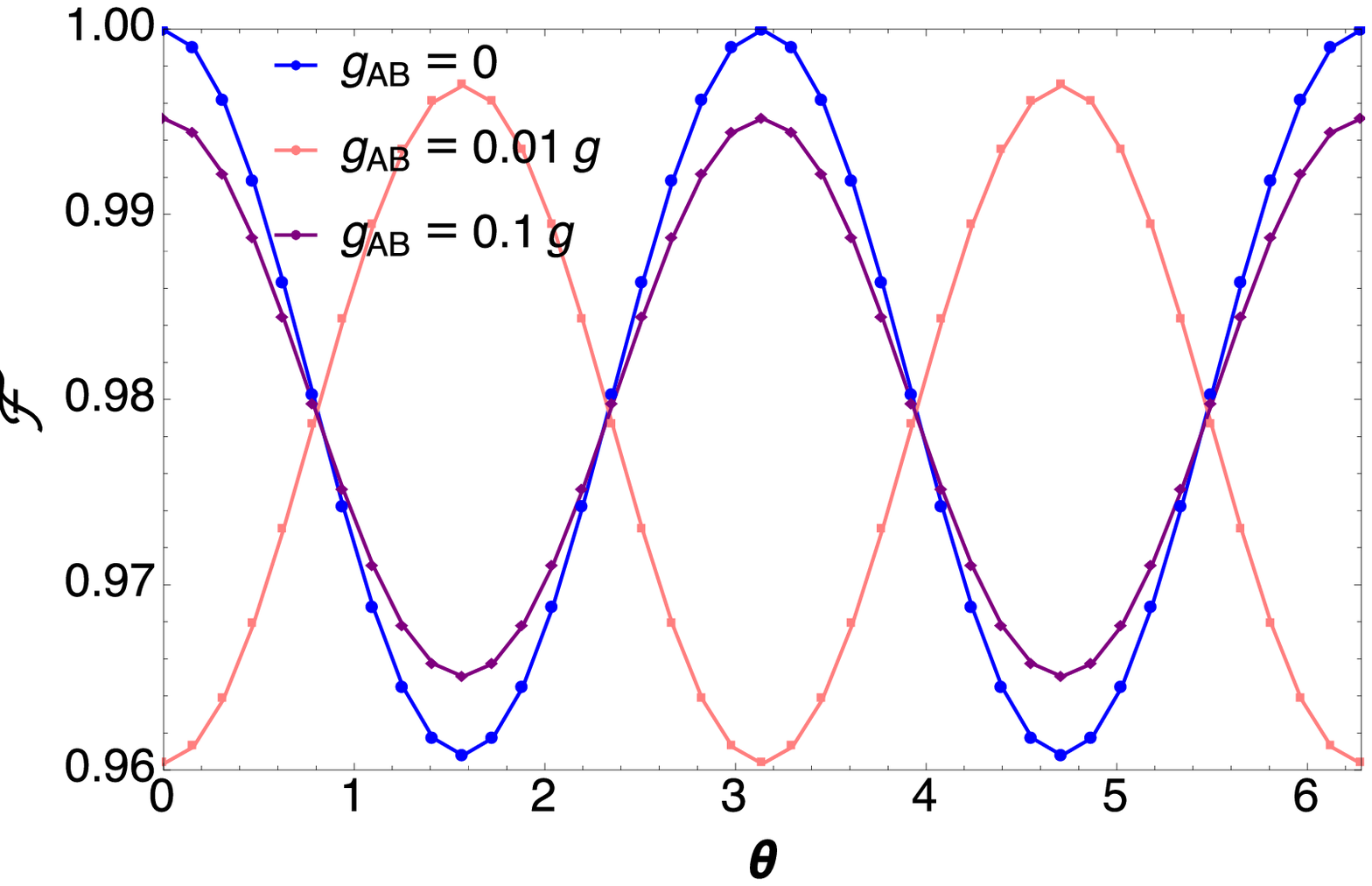}}
\subfigure[]{\includegraphics[width=3in]{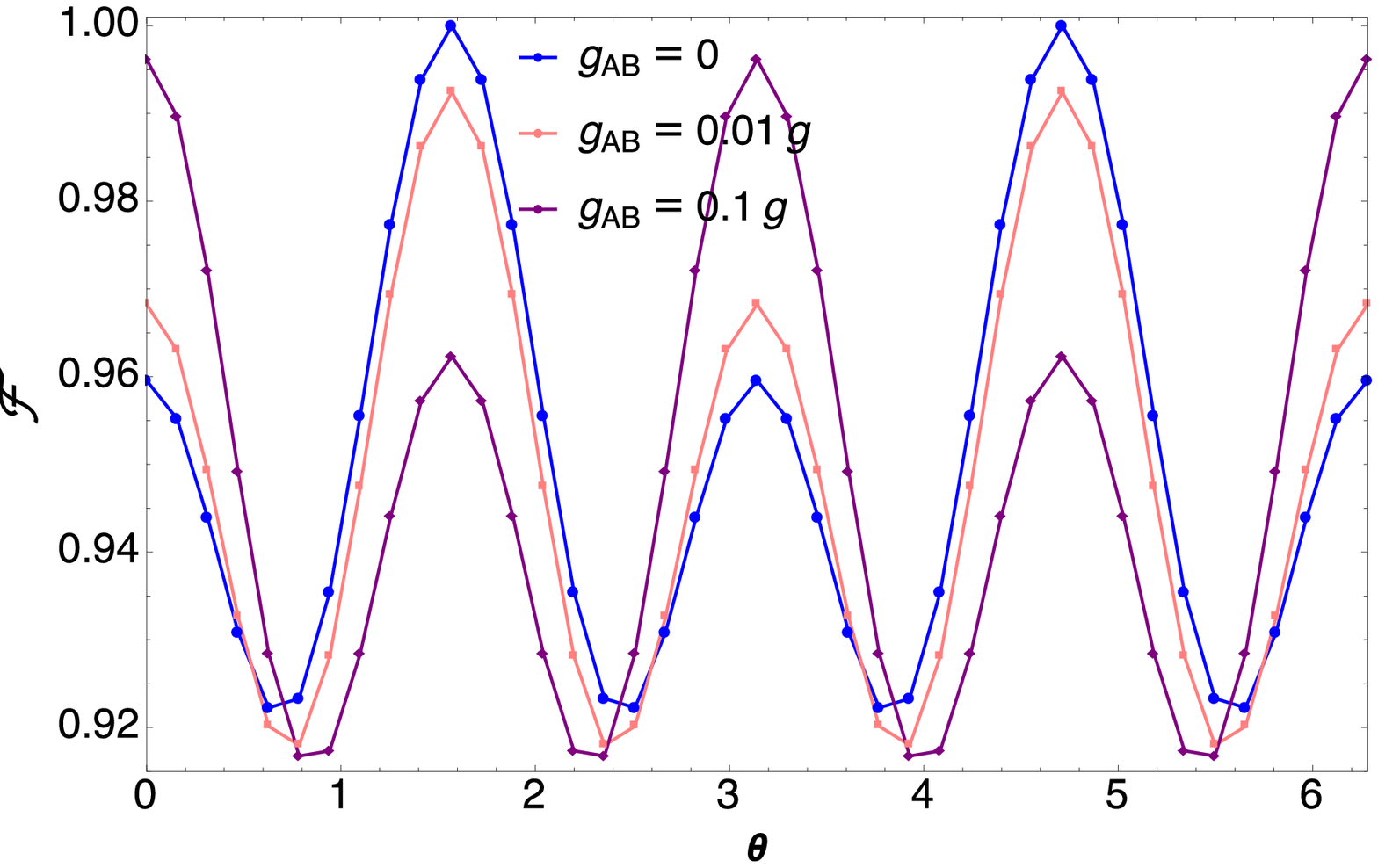}}
\end{center}
\caption{(color online) Fidelity $\mathcal{F}$ versus $\theta$  and by taking the unwanted inter-cavity crosstalk into account for $g_{AB}=0,~0.01g,~0.1g$. }
\label{fig:}
\end{figure}
Figure~8 (a) (b) display the effect of the unwanted inter-cavity crosstalk on the fidelity, respectively.
The effect of the inter-cavity crosstalk can be taken into 
account by adding a Hamiltonian of the form $H_{AB}=g_{AB}(a^{\dagger }b+ab^{\dagger })$ in $H_I$,
where $g_{AB}$ is the inter-cavity crosstalk coupling strength. Figures 8 (a) and (b) display fidelity $\mathcal{F}$ versus $\theta$ for $g_{AB}=0,~0.01g,~0.1g$, which consider the case of  (\romannumeral1) the state $|g\rangle$ acts as a control state and
(\romannumeral2) the state $|e\rangle$ acts as a control state, respectively.
From the Fig.~8 (a) and (b)  we
can see that the effect of the crosstalk on the fidelity is negligibly small  for $g_{AB}=0.01g,~0.1g$.
Moreover, we calculates the average fidelities are approximately (\romannumeral1) $98.13\%$, $97.82\%$, and $98.03\%$, (\romannumeral2) $95.21\%$, $95.03\%$, and $95.05\%$ for $g_{AB}=0,~0.01g,~0.1g$, respectively. 
In Fig.~9 (a) and 9 (b), we choose  the inter-cavity crosstalk coupling strengths
$g_{AB}=0.1g$ and $g_{AB}=0.01g$, respectively.

\begin{figure}[tbp]
\begin{center}
\subfigure[]{\includegraphics[width=3.5in]{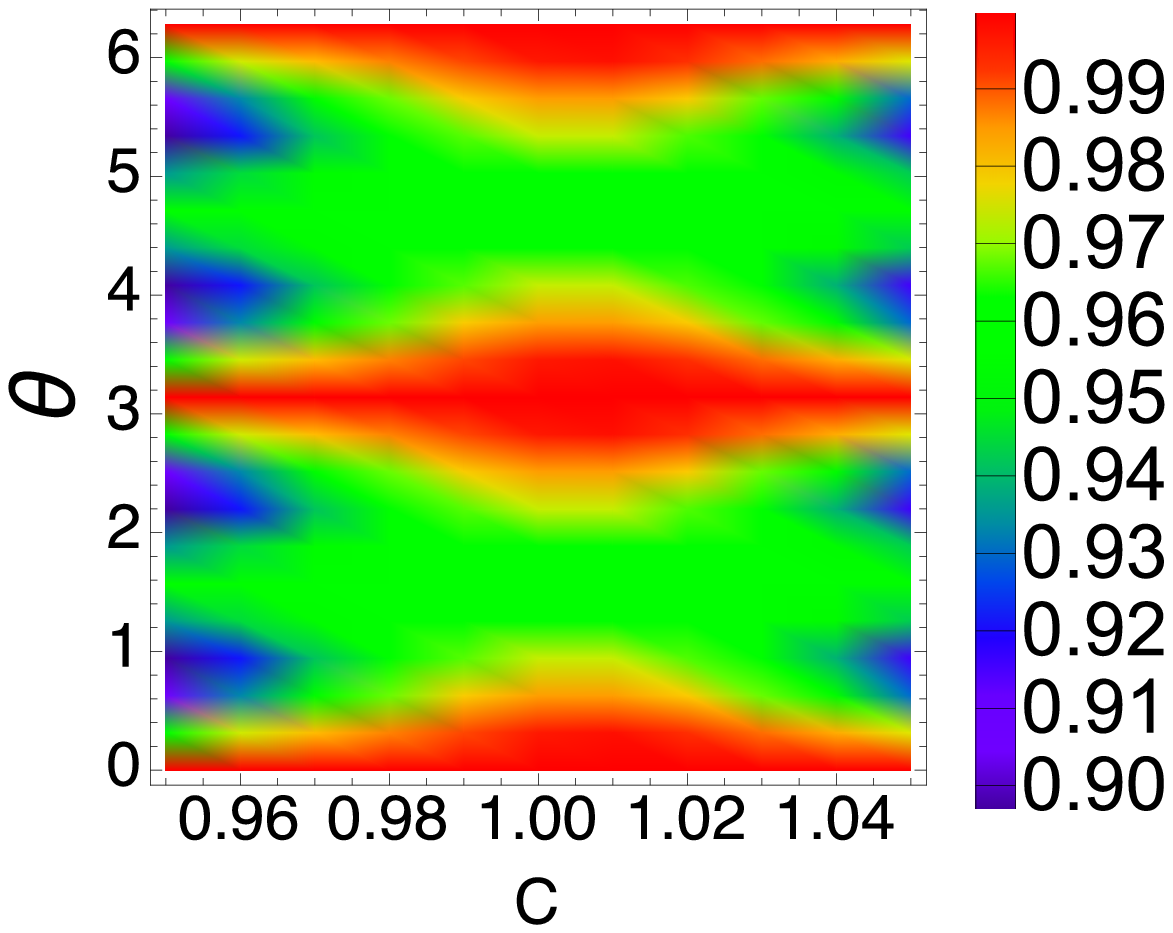}}
\subfigure[]{\includegraphics[width=3.5in]{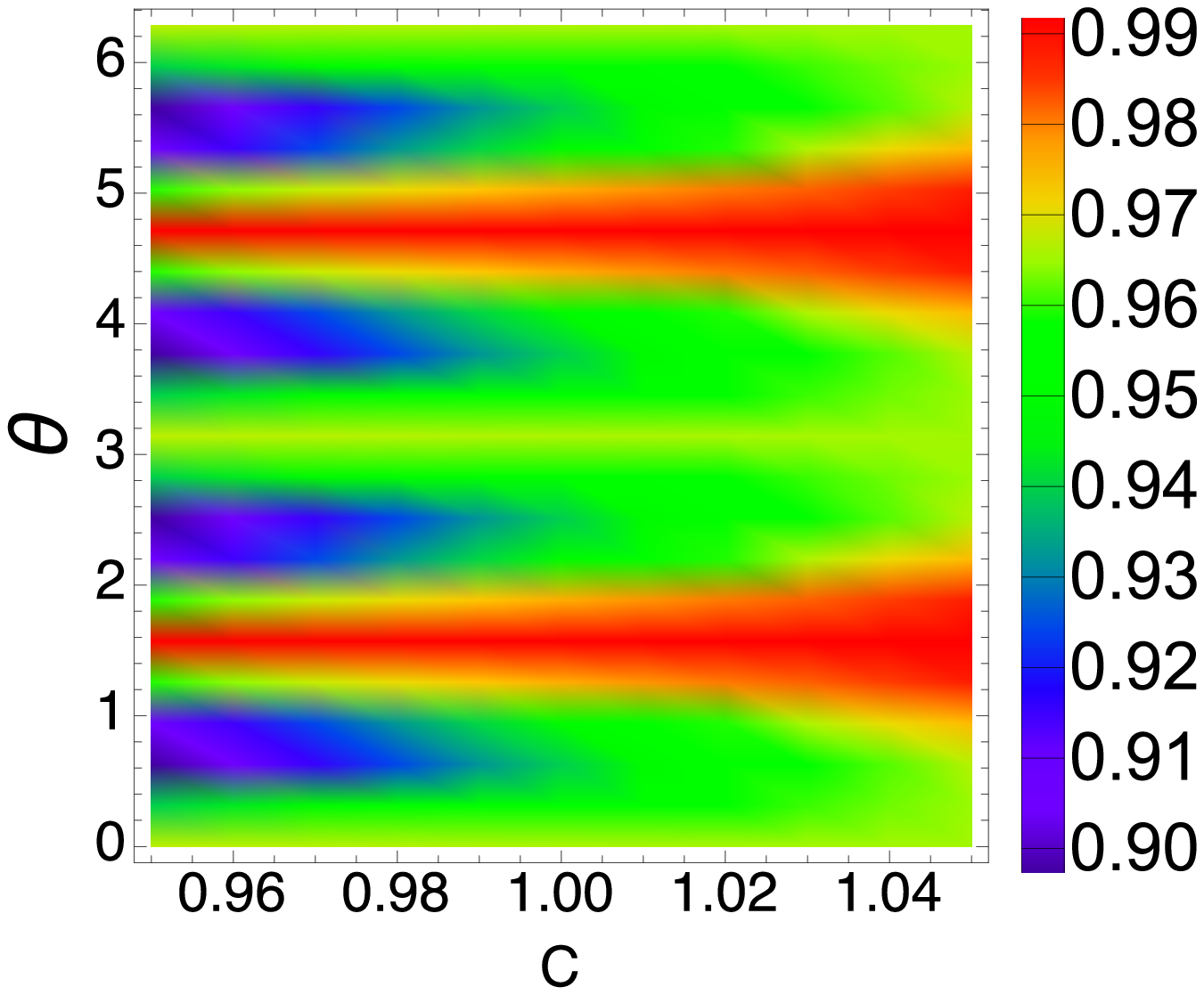}}
\end{center}
\caption{color online) Fidelity $\mathcal{F}$ versus $c$ and $\theta$ by taking into account the inhomogeneity in transmon-cavity interaction. Here $g_A=g$ and $g_B=cg$ with $c\in[0.95,1.05]$.}
\label{fig:}
\end{figure}

Figure 9 (a) and 9 (b) take into account the inhomogeneity in transmon-cavity interaction for 
the case of  (\romannumeral1) the state $|g\rangle$ acts as a control state and
(\romannumeral2) the state $|e\rangle$ acts as a control state, respectively. Figure~9 displays the fidelity versus $c$ and $\theta$, where we set
$g_A=g$ and $g_B=cg$ with $c\in[0.95,1.05]$. Figure 9 (a) shows that for  $0.98\leq c <1$ and $1< c\leq 1.02$, the effect of the inhomogeneity on the fidelity is very small. Moreover, one can see that for $0.97\leq c \leq1.03$, the fidelity can be greater than 95.61\%. Figure 9 (b) displays that the fidelity is almost unaffected by the inhomogeneity for $0.99\leq c \leq1.05$. 
It is interesting to note that the optimum fidelity has a small increase in the region  $1< c \leq1.05$.

The simulations above show that the quantum adder that generates a superposition of two unknown states
with high fidelity can be
achieved for small errors in the undesired the weak anharmonicity, inter-cavity crosstalk, and the inhomogeneity.  

\section{Conclusion}
\label{sec6}
We have devised a quantum adder based on 3D circuit QED that is a good candidate for quantum information processing, computation and simulation. The 3D microwave cavities dispersively interacting with the transmon to effectively create a superposition state of two arbitrary states
encoded in two cavities. 
The initial state of each cavity is arbitrarily that selected as the discrete-variable
state or the continuous-variable state.
Our proposal also can be applied to
other types of qubits such as natural atoms \cite{05atom} and artificial atoms (other superconducting qubits (e.g., phase qubits \cite{11phase}, Xmon
qubits \cite{13Xmon}, flux qubit \cite{16flux}), NV centers \cite{11zhu}, and quantum dots \cite{07dot}).
The Numerical
simulations imply that
the high-fidelity generation of a superposition state of two  cavities is
feasible with the current circuit QED technology.
Finally, our proposal
provides a way for realizing a quantum adder in a 3D circuit
QED system, and such quantum machine may find other 
applications in quantum information processing and
computation.
We hope this work would stimulate experimental activities in the near future.

\section*{ACKNOWLEDGEMENTS}
This work was supported by the
National Natural Science Foundation of China, under Grant No.11775040, 
No. 11375036, and No.11847128, and the Fundamental
Research Fund for the Central Universities under Grants No. DUT18LK45.

\appendix

\begin{widetext}
\section{Derivation of the equations~(\ref{eq109}) and~(\ref{eq113})}
\label{Appendixb}
Based on Hamiltonians
\begin{eqnarray} \label{ad3}
\begin{split}
 H_{I}^{g}=-\chi(a^{\dagger }b+a b^{\dagger })\\
H_{I}^{f}=\lambda\left(a^\dagger b +a
b^\dagger \right),
\end{split}
\end{eqnarray}
we can obtain the following transformations
\begin{eqnarray}\label{ad4}
a^{\dagger }(t)&=&\cos (\chi t)a^{\dagger }+i\sin (\chi t)b^\dagger,     \nonumber \\
b^\dagger(t)&=&\cos (\chi
t)b^{\dagger }+i\sin (\chi t)a^\dagger,
\end{eqnarray}
for $|g\rangle$ component and
\begin{eqnarray}\label{ad5}
a^{\dagger }(t)&=&\cos (\lambda t)a^{\dagger }-i\sin (\lambda t)b^\dagger, \nonumber \\
b^\dagger(t)&=&\cos (\lambda
t)b^{\dagger }-i\sin (\lambda t)a^\dagger,
\end{eqnarray}
for $|f\rangle$ component, respectively. 

\textit{(\romannumeral1) $|g\rangle$ acts as a control state.   }
When the evolution time $t=\pm(\frac{\pi}{2}+2k_1\pi)/\chi$,
one has $a^{\dagger }(t)=ib^\dagger$ and $b^\dagger(t)=ia^\dagger$ for Eq.~(\ref{ad4}). 
When the condition $\lambda
t=\pm(2\pi+2k_2\pi)$ is satisfied, we also have
$a^{\dagger }(t)=a^\dagger$ and $b^\dagger(t)=b^\dagger$
for Eq.~(\ref{ad5}). Here, the sign $``+"$ or $``-"$ depends on the detuning $\Delta>0$ or $\Delta<0$, and $k_1$ and $k_2$ are  non-negative integers. 

We suppose $t=\pm(\frac{\pi}{2}+2k_1\pi)/\chi$ and $\lambda
t=\pm(2\pi+2k_2\pi)$,  the state~(\ref{eq106}) 
\begin{eqnarray} \label{ad6}
\sin\theta\exp{\left(-iH_{0}^{g}t\right)}\exp{(-iH_{I}^{g}t)}|g\rangle|\psi \rangle _{A}|\varphi \rangle
_{B} +\cos\theta\exp{\left(-iH_{0}^{f}t\right) }\exp{\left( -iH_{I}^{f}t\right) }|f\rangle |\psi \rangle _{A}|\varphi \rangle 
_{B} 
\end{eqnarray}
of the system changes as
\begin{eqnarray}\label{add1}
&&\sin\theta\exp{\left(-iH_{0}^{g}t\right)}\exp{(-iH_{I}^{g}t)}|g\rangle|\psi \rangle _{A}|\varphi \rangle
_{B} +\cos\theta\exp{\left(-iH_{0}^{f}t\right) }\exp{\left( -iH_{I}^{f}t\right) }|f\rangle |\psi \rangle _{A}|\varphi \rangle 
_{B}  \nonumber \\
&=&\sin\theta\exp{\left(-iH_{0}^{g}t\right)}\exp{(-iH_{I}^{g}t)}|g\rangle \sum\limits_{n=0}^{\infty }\frac{c_{n}}{\sqrt{n!}}(a^{\dagger})^{n}|0\rangle_{A}\sum\limits_{m=0}^{\infty }\frac{d_{m}}{\sqrt{m!}}(b^{\dagger})^{m}|0\rangle_{B}  \nonumber\\
&+&\cos\theta\exp{\left(-iH_{0}^{f}t\right) }\exp{\left( -iH_{I}^{f}t\right) }|f\rangle  \sum\limits_{n=0}^{\infty }\frac{c_{n}}{\sqrt{n!}}(a^{\dagger})^{n}|0\rangle_{A}\sum\limits_{m=0}^{\infty }\frac{d_{m}}{\sqrt{m!}}(b^{\dagger})^{m}|0\rangle_{B} \nonumber \\
&=&\sin\exp{\left(-iH_{0}^{g}t\right)}|g\rangle \sum\limits_{n=0}^{\infty }\frac{c_{n}}{\sqrt{n!}}(ib^{\dagger})^{n}|0\rangle_{B}\sum\limits_{m=0}^{\infty }\frac{d_{m}}{\sqrt{m!}} (ia^{\dagger})^{m}|0\rangle_{A} \nonumber \\
&+&\cos\theta \exp{\left(-iH_{0}^{f}t\right) }|f\rangle  \sum\limits_{n=0}^{\infty }\frac{c_{n}}{\sqrt{n!}}(a^{\dagger})^{n}|0\rangle_{A}\sum\limits_{m=0}^{\infty }\frac{d_{m}}{\sqrt{m!}}(b^{\dagger})^{m}|0\rangle_{B} \nonumber \\
&=&\sin\theta \exp{\left(-iH_{0}^{g}t\right)}|g\rangle \sum\limits_{n=0}^{\infty }(i)^{n}c_{n}|n\rangle _{B}\sum\limits_{m=0}^{\infty }(i)^{m}d_{m}|m\rangle _{A}
+\cos\theta \exp{\left(-iH_{0}^{f}t\right) }|f\rangle  \sum\limits_{n=0}^{\infty }c_{n}|n\rangle _{A}\sum\limits_{m=0}^{\infty }d_{m}|m\rangle _{B}\nonumber 
\end{eqnarray}
\begin{eqnarray}
&=&\sin\theta \exp{\left(i\chi mt\right)}\exp{\left(i\chi nt\right)}\exp{\left(i \frac{\pi}{2} m\right)}\exp{\left(i \frac{\pi}{2} n\right)}|g\rangle |\psi \rangle _{B}|\varphi \rangle
_{A} \nonumber \\
&+&\cos\theta \exp{\left(-i\lambda nt\right)}\exp{\left(-i\lambda mt\right)}\exp{\left(-i2\lambda t\right)}|f\rangle |\psi\rangle_A|\varphi\rangle_B \nonumber \\
&=&\sin\theta \exp{\left(i m\pi[\pm(\frac{1}{2}+2k_1)+\frac{1}{2}]\right)}\exp{\left(i n\pi[\pm(\frac{1}{2}+2k_1)+\frac{1}{2}\right)} |g\rangle |\varphi \rangle
_{A}|\psi \rangle _{B}\nonumber \\
&+&\cos\theta \exp{\left[\mp in\pi(2+2k_2)\right]}\exp{\left[\mp im\pi(2+2k_2)\right]}\exp{\left[\mp i2\pi(2+2k_2)\right]}
|f\rangle |\psi\rangle_A|\varphi\rangle_B\nonumber \\
&=&\sin\theta|g\rangle |\varphi \rangle
_{A}|\psi \rangle _{B}
+\cos\theta|f\rangle |\psi\rangle_A|\varphi\rangle_B,\label{eq:er2}
\end{eqnarray}
where we have used $a^{\dagger }a|n\rangle =n|n\rangle$, $b^{\dagger
}b|m\rangle =m|m\rangle$,
$(i)^{n}=e^{i \frac{\pi}{2} n}$, and $(i)^{m}=e^{i \frac{\pi}{2} m}$. Here that $ |\psi\rangle$ and $|\varphi\rangle$ have the same Hilbert space while they are arbitrary asymmetric states.

\textit{(\romannumeral2) $|f\rangle$ acts as a control state.} For the evolution time $t=\pm(\frac{\pi}{2}+2k_1\pi)/\lambda$,
one has $a^{\dagger }(t)=-ib^\dagger$ and $b^\dagger(t)=-ia^\dagger$ for the level $|f\rangle$. For the condition $\chi
t=\pm(2\pi+2k_2\pi)$, one also has
$a^{\dagger }(t)=a^\dagger$ and $b^\dagger(t)=b^\dagger$
for the level $|g\rangle$. Here, the sign $``+" $ or $``-"$ depends on the detuning $\Delta>0$ or $\Delta<0$, and $k_1$ and $k_2$ are  non-negative integers. 

Under the conditions of $t=\pm(\frac{\pi}{2}+2k_1\pi)/\lambda$ and $\chi
t=\pm(2\pi+2k_2\pi)$, the system state~(\ref{eq106}) 
\begin{eqnarray} \label{ad7}
\sin\theta\exp{\left(-iH_{0}^{g}t\right)}\exp{(-iH_{I}^{g}t)}|g\rangle|\psi \rangle _{A}|\varphi \rangle
_{B} +\cos\theta\exp{\left(-iH_{0}^{f}t\right) }\exp{\left( -iH_{I}^{f}t\right) }|f\rangle |\psi \rangle _{A}|\varphi \rangle 
_{B} 
\end{eqnarray}
becomes 
\begin{eqnarray}\label{eq:er3}
&&\sin\theta\exp{\left(-iH_{0}^{g}t\right)}\exp{(-iH_{I}^{g}t)}|g\rangle|\psi \rangle _{A}|\varphi \rangle
_{B} +\cos\theta\exp{\left(-iH_{0}^{f}t\right) }\exp{\left( -iH_{I}^{f}t\right) }|f\rangle |\psi \rangle _{A}|\varphi \rangle 
_{B}  \nonumber \\
&=&\sin\theta\exp{\left(-iH_{0}^{g}t\right)}\exp{(-iH_{I}^{g}t)}|g\rangle \sum\limits_{n=0}^{\infty }\frac{c_{n}}{\sqrt{n!}}(a^{\dagger})^{n}|0\rangle_{A}\sum\limits_{m=0}^{\infty }\frac{d_{m}}{\sqrt{m!}}(b^{\dagger})^{m}|0\rangle_{B}  \nonumber \\
&+&\cos\theta\exp{\left(-iH_{0}^{f}t\right) }\exp{\left( -iH_{I}^{f}t\right) }|f\rangle  \sum\limits_{n=0}^{\infty }\frac{c_{n}}{\sqrt{n!}}(a^{\dagger})^{n}|0\rangle_{A}\sum\limits_{m=0}^{\infty }\frac{d_{m}}{\sqrt{m!}}(b^{\dagger})^{m}|0\rangle_{B} \nonumber \\
&=&\sin\theta\exp{\left(-iH_{0}^{g}t\right)}|g\rangle \sum\limits_{n=0}^{\infty }\frac{c_{n}}{\sqrt{n!}}(a^{\dagger})^{n}|0\rangle_{A}\sum\limits_{m=0}^{\infty }\frac{d_{m}}{\sqrt{m!}}(b^{\dagger})^{m}|0\rangle_{B} \nonumber \\
&+&\cos\theta\exp{\left(-iH_{0}^{f}t\right) }|f\rangle  \sum\limits_{n=0}^{\infty }\frac{c_{n}}{\sqrt{n!}}(-ib^{\dagger})^{n}|0\rangle_{B}\sum\limits_{m=0}^{\infty }\frac{d_{m}}{\sqrt{m!}} (-ia^{\dagger})^{m}|0\rangle_{A} \nonumber \\
&=&\sin\theta \exp{\left(-iH_{0}^{g}t\right)}|g\rangle \sum\limits_{n=0}^{\infty }c_{n}|n\rangle _{A}\sum\limits_{m=0}^{\infty }d_{m}|m\rangle _{B} 
+\cos\theta \exp{\left(-iH_{0}^{f}t\right) }|f\rangle  \sum\limits_{n=0}^{\infty }(-i)^{n}c_{n}|n\rangle _{B}\sum\limits_{m=0}^{\infty }(-i)^{m}d_{m}|m\rangle _{A} \nonumber \\
&=&\sin\theta \exp{\left(i\chi mt\right)}\exp{\left(i\chi nt\right)} |g\rangle |\psi \rangle _{A}|\varphi \rangle
_{B} \nonumber \\
&+&\cos\theta \exp{\left(-i\lambda mt\right)} \exp{\left(-i\lambda nt\right)} \exp{\left(-i2\lambda t\right)} \exp{\left(-i \frac{\pi}{2} m\right)} \exp{\left(-i \frac{\pi}{2} n\right)} 
|f\rangle |\psi\rangle_B|\varphi\rangle_A \nonumber \\
&=&\sin\theta \exp{\left[\pm i2m\pi(1+k_2)\right]} \exp{\left[\pm i2n\pi(1+k_2)\right]} 
|g\rangle |\psi \rangle _{A}|\varphi \rangle
_{B} \nonumber \\
&+&\cos\theta \exp{\left(i m\pi[\pm(\frac{1}{2}+2k_1)-\frac{1}{2}]\right)}\exp{\left(i n\pi[\pm(\frac{1}{2}+2k_1)-\frac{1}{2}]\right)}\exp{\left[\mp i\pi(1+4k_1)\right]}
|f\rangle  |\varphi\rangle_A |\psi\rangle_B\nonumber \\
&=&\sin\theta|g\rangle  |\psi\rangle_A|\varphi\rangle_B 
-\cos\theta|f\rangle |\varphi \rangle
_{A}|\psi \rangle _{B},
\end{eqnarray}
where we have used $(-i)^{n}=e^{-i \frac{\pi}{2} n}$, and $(-i)^{m}=e^{-i \frac{\pi}{2} m}$. It should be noted here that $ |\psi\rangle$ and $|\varphi\rangle$ have the same Hilbert space while they are arbitrary asymmetric states.

\section{Derivation of the equations~(\ref{eq118}) and~(\ref{eq133})}
\label{Appendixc}

By solving the Heisenberg equations for 
\begin{eqnarray} \label{ad8}
\begin{split}
H_{I}^{g}=-\chi(a^{\dagger }b+a b^{\dagger }),\\
H_{I}^{e}=\Lambda\left(a^\dagger b +a
b^\dagger \right),
\end{split}
\end{eqnarray}
the dynamics of the operators $a^{\dagger }$ and $b^{\dagger }$ can be derived
as
\begin{eqnarray}\label{ad9}
a^{\dagger }(t)&=&\cos (\chi t)a^{\dagger }+i\sin (\chi t)b^\dagger,     \nonumber \\
b^\dagger(t)&=&\cos (\chi
t)b^{\dagger }+i\sin (\chi t)a^\dagger, 
\end{eqnarray}
for $|g\rangle$ component, and 
\begin{eqnarray}\label{ad10}
a^{\dagger }(t)&=&\cos (\Lambda t)a^{\dagger }-i\sin (\Lambda t)b^\dagger, \nonumber \\
b^\dagger(t)&=&\cos (\Lambda
t)b^{\dagger }-i\sin (\Lambda t)a^\dagger,
\end{eqnarray}
for $|e\rangle$ component.

\textit{(\romannumeral1) $|g\rangle$ acts as a control state.} When the evolution time $t=\pm(\frac{\pi}{2}+2k_1\pi)/\chi$ and the condition $\Lambda
t=\mp(2\pi+2k_2\pi)$,
we have $a^{\dagger }(t)=ib^\dagger$ and $b^\dagger(t)=ia^\dagger$ for Eq.~(\ref{ad9}) and $a^{\dagger }(t)=a^\dagger$ and $b^\dagger(t)=b^\dagger$
for Eq.~(\ref{ad10}), respectively. Here, the sign $``+"$ or $``-"$ of the expression $t$ depends on the detuning $\Delta>0$ or $\Delta<0$, and $k_1$ and $k_2$ are  non-negative integers. 
Under the conditions of $t=\pm(\frac{\pi}{2}+2k_1\pi)/\chi$ and $\Lambda
t=\mp(2\pi+2k_2\pi)$,
the state~(\ref{eq115}) 
\begin{eqnarray}\label{ad11}
\sin\theta\exp{\left(-iH_{0}^{g}t\right)}\exp{(-iH_{I}^{g}t)}|g\rangle|\psi \rangle _{A}|\varphi \rangle
_{B} +\cos\theta\exp{\left(-iH_{0}^{e}t\right) }\exp{\left( -iH_{I}^{e}t\right) }|e\rangle |\psi \rangle _{A}|\varphi \rangle 
_{B}
\end{eqnarray}
changes to
\begin{eqnarray}\label{eq:er4}
\begin{split}
&\sin\theta\exp{\left(-iH_{0}^{g}t\right)}\exp{(-iH_{I}^{g}t)}|g\rangle|\psi \rangle _{A}|\varphi \rangle
_{B} +\cos\theta\exp{\left(-iH_{0}^{e}t\right) }\exp{\left( -iH_{I}^{e}t\right) }|e\rangle |\psi \rangle _{A}|\varphi \rangle 
_{B}  \\
=&\sin\theta \exp{\left(-iH_{0}^{g}t\right)}\exp{(-iH_{I}^{g}t)}|g\rangle \sum\limits_{n=0}^{\infty }\frac{c_{n}}{\sqrt{n!}}(a^{\dagger})^{n}|0\rangle_{A}\sum\limits_{m=0}^{\infty }\frac{d_{m}}{\sqrt{m!}}(b^{\dagger})^{m}|0\rangle_{B}  \\
+&\cos\theta \exp{\left(-iH_{0}^{e}t\right) }\exp{\left( -iH_{I}^{e}t\right) }|e\rangle  \sum\limits_{n=0}^{\infty }\frac{c_{n}}{\sqrt{n!}}(a^{\dagger})^{n}|0\rangle_{A}\sum\limits_{m=0}^{\infty }\frac{d_{m}}{\sqrt{m!}}(b^{\dagger})^{m}|0\rangle_{B}  \\
=&\sin\theta \exp{\left(-iH_{0}^{g}t\right)}|g\rangle \sum\limits_{n=0}^{\infty }\frac{c_{n}}{\sqrt{n!}}(ib^{\dagger})^{n}|0\rangle_{B}\sum\limits_{m=0}^{\infty }\frac{d_{m}}{\sqrt{m!}} (ia^{\dagger})^{m}|0\rangle_{A} \\
+&\cos\theta \exp{\left(-iH_{0}^{e}t\right) }|e\rangle  \sum\limits_{n=0}^{\infty }\frac{c_{n}}{\sqrt{n!}}(a^{\dagger})^{n}|0\rangle_{A}\sum\limits_{m=0}^{\infty }\frac{d_{m}}{\sqrt{m!}}(b^{\dagger})^{m}|0\rangle_{B} \\
=&\sin\theta  \exp{\left(-iH_{0}^{g}t\right)}g\rangle \sum\limits_{n=0}^{\infty }(i)^{n}c_{n}|n\rangle _{B}\sum\limits_{m=0}^{\infty }(i)^{m}d_{m}|m\rangle _{A} 
+\cos\theta \exp{\left(-iH_{0}^{e}t\right) }|e\rangle  \sum\limits_{n=0}^{\infty }c_{n}|n\rangle _{A}\sum\limits_{m=0}^{\infty }d_{m}|m\rangle _{B} \\
=&\sin\theta  \exp{\left(i\chi mt\right)} \exp{\left(i\chi nt\right)} \exp{\left(i \frac{\pi}{2} m\right)}
\exp{\left(i \frac{\pi}{2} n\right)} 
|g\rangle |\psi \rangle _{B}|\varphi \rangle
_{A} \\
+&\cos\theta \exp{\left(-i\Lambda nt\right)} \exp{\left(-i\Lambda mt\right)} \exp{\left(-i2\chi t\right)} 
|e\rangle |\psi\rangle_A|\varphi\rangle_B\\
=&\sin\theta  \exp{\left(i m\pi[\pm(\frac{1}{2}+2k_1)+\frac{1}{2}]\right)} \exp{\left(i n\pi[\pm(\frac{1}{2}+2k_1)+\frac{1}{2}]\right)}
|g\rangle |\varphi \rangle
_{A}|\psi \rangle _{B} \\
+&\cos\theta \exp{\left[\pm i2n\pi(1+k_2)\right]} \exp{\left[\pm i2m\pi(1+k_2)\right]} \exp{\left(\mp i2\pi(\frac{1}{2}+2k_1)\right)} 
|e\rangle |\psi\rangle_A|\varphi\rangle_B\\
=&\sin\theta|g\rangle |\varphi \rangle
_{A}|\psi \rangle _{B}
-\cos\theta|e\rangle |\psi\rangle_A|\varphi\rangle_B.
\end{split}
\end{eqnarray}
where we have used $(i)^{n}=e^{i \frac{\pi}{2} n}$, and $(i)^{m}=e^{i \frac{\pi}{2} m}$. Here, the states $ |\psi\rangle$ and $|\varphi\rangle$ have the same Hilbert space while they are arbitrary asymmetric states.

\textit{(\romannumeral2) $|e\rangle$ acts as a control state.} For the evolution time $t=\mp(\frac{\pi}{2}+2k_1\pi)/\Lambda$ and the condition $\chi
t=\pm(2\pi+2k_2\pi)$,
one has $a^{\dagger }(t)=-ib^\dagger$ and $b^\dagger(t)=-ia^\dagger$ for Eq.~(\ref{ad9}) , and $a^{\dagger }(t)=a^\dagger$ and $b^\dagger(t)=b^\dagger$ for Eq.~(\ref{ad10}), respectively. 
 Here the sign $``-"$ of the expression $t$
corresponds to $\Delta>0$ while the sign $``+"$ corresponds to $\Delta<0$. 
When the conditions $t=(\frac{\pi}{2}+2k_1\pi)/|\Lambda|$ and 
$|\chi|t=2\pi+2k_2\pi$ are satisfied, the Eq.~(\ref{eq115})
\begin{eqnarray}\label{ad12}
\sin\theta\exp{\left(-iH_{0}^{g}t\right)}\exp{(-iH_{I}^{g}t)}|g\rangle|\psi \rangle _{A}|\varphi \rangle
_{B} +\cos\theta\exp{\left(-iH_{0}^{e}t\right) }\exp{\left( -iH_{I}^{e}t\right) }|e\rangle |\psi \rangle _{A}|\varphi \rangle 
_{B}
\end{eqnarray}
changes to
\begin{eqnarray}\label{eq:er5}
&&\sin\theta\exp{\left(-iH_{0}^{g}t\right)}\exp{(-iH_{I}^{g}t)}|g\rangle|\psi \rangle _{A}|\varphi \rangle
_{B} +\cos\theta\exp{\left(-iH_{0}^{e}t\right) }\exp{\left( -iH_{I}^{e}t\right) }|e\rangle |\psi \rangle _{A}|\varphi \rangle 
_{B} \nonumber \\
&=&\sin\theta \exp{\left(-iH_{0}^{g}t\right)}\exp{(-iH_{I}^{g}t)}|g\rangle \sum\limits_{n=0}^{\infty }\frac{c_{n}}{\sqrt{n!}}(a^{\dagger})^{n}|0\rangle_{A}\sum\limits_{m=0}^{\infty }\frac{d_{m}}{\sqrt{m!}}(b^{\dagger})^{m}|0\rangle_{B}  \nonumber \\
&+&\cos\theta \exp{\left(-iH_{0}^{e}t\right) }\exp{\left( -iH_{I}^{e}t\right) }|e\rangle  \sum\limits_{n=0}^{\infty }\frac{c_{n}}{\sqrt{n!}}(a^{\dagger})^{n}|0\rangle_{A}\sum\limits_{m=0}^{\infty }\frac{d_{m}}{\sqrt{m!}}(b^{\dagger})^{m}|0\rangle_{B}  \nonumber \\
&=&\sin\theta \exp{\left(-iH_{0}^{g}t\right)}|g\rangle \sum\limits_{n=0}^{\infty }\frac{c_{n}}{\sqrt{n!}}(a^{\dagger})^{n}|0\rangle_{A}\sum\limits_{m=0}^{\infty }\frac{d_{m}}{\sqrt{m!}}(b^{\dagger})^{m}|0\rangle_{B}\nonumber \\
&+&\cos\theta \exp{\left(-iH_{0}^{e}t\right) }|e\rangle  
 \sum\limits_{n=0}^{\infty }\frac{c_{n}}{\sqrt{n!}}(-ib^{\dagger})^{n}|0\rangle_{B}\sum\limits_{m=0}^{\infty }\frac{d_{m}}{\sqrt{m!}} (-ia^{\dagger})^{m}|0\rangle_{A} \nonumber \\
&=&\sin\theta \exp{\left(-iH_{0}^{g}t\right)}|g\rangle  \sum\limits_{n=0}^{\infty }c_{n}|n\rangle _{A}\sum\limits_{m=0}^{\infty }d_{m}|m\rangle _{B} 
+\cos\theta \exp{\left(-iH_{0}^{e}t\right) }|e\rangle \sum\limits_{n=0}^{\infty }(-i)^{n}c_{n}|n\rangle _{B}\sum\limits_{m=0}^{\infty }(-i)^{m}d_{m}|m\rangle _{A} \nonumber \\
&=&\sin\theta \exp{\left(i\chi mt\right)} \exp{\left(i\chi nt\right)} 
|g\rangle |\psi \rangle _{A}|\varphi \rangle
_{B} \nonumber \\
&+&\cos\theta \exp{\left(-i\Lambda nt\right)} \exp{\left(-i\Lambda mt\right)} \exp{\left(-i2\chi t\right)} \exp{\left(-i \frac{\pi}{2} m\right)}
\exp{\left(-i \frac{\pi}{2} n\right)} 
|e\rangle |\psi\rangle_B|\varphi\rangle_A\nonumber \\
&=&\sin\theta  \exp{\left[\pm i2n\pi(1+k_2)\right]} \exp{\left[\pm i2m\pi(1+k_2)\right]} 
|g\rangle |\psi \rangle _{A} |\varphi \rangle
_{B}\nonumber \\
&+&\cos\theta  \exp{\left(i m\pi[\pm(\frac{1}{2}+2k_1)-\frac{1}{2}]\right)} \exp{\left(i n\pi[\pm(\frac{1}{2}+2k_1)-\frac{1}{2}]\right)}\exp{\left[\mp i4\pi(1+k_2)\right]}  |e\rangle |\psi\rangle_A|\varphi\rangle_B \nonumber \\
&=&\sin\theta |g\rangle|\psi\rangle_A|\psi \rangle _{B}
+\cos\theta|e\rangle  |\varphi \rangle
_{A}|\psi \rangle _{B},
\end{eqnarray}
where we have used $(-i)^{n}=e^{-i \frac{\pi}{2} n}$, and $(-i)^{m}=e^{-i \frac{\pi}{2} m}$. Here, the states $ |\psi\rangle$ and $|\varphi\rangle$ have the same Hilbert space while they are arbitrary asymmetric states.

\end{widetext}


\begin{thebibliography}{99}

\bibitem{09entanglement} R. Horodecki, P. Horodecki, M. Horodecki, and K. Horodecki, Quantum entanglement, Rev. Mod. Phys. \textbf{81}, 865 (2009).

\bibitem{14coherence} T. Baumgratz, M. Cramer, and M. B. Plenio, Quantifying coherence, Phys. Rev. Lett. \textbf{113}, 140401 (2014).

\bibitem{17coherence} C. S. Yu, Quantum coherence via skew information and its polygamy,
Phys. Rev. A \textbf{95}, 042337 (2017).

\bibitem{97Shor} P. W. Shor, Polynomial-time algorithms for prime factorization and discrete logarithms on a quantum computer, SIAM J.
Comput. \textbf{26}, 1484 (1997).

\bibitem{96Grover} L. K. Grover, A fast quantum mechanical algorithm for database
search, in \emph{Proceedings of the Twenty-Eighth Annual ACM
Symposium on Theory of Computing} (ACM Press, New York,
1996), pp. 212-219.

\bibitem{11Giovannetti} V. Giovannetti, S. Lloyd, and L. Maccone, Advances in quantum metrology, Nat. Photonics \textbf{5},
222 (2011).

\bibitem{02cryptography} N. Gisin, G. Ribordy, W. Tittel, and H. Zbinden, Quantum cryptography, Rev. Mod. Phys. \textbf{74}, 745 (2002).

\bibitem{15adder} Y. Alvarez-Rodriguez, M. Sanz, L. Lamata, and E. Solano,
The forbidden quantum adder, Sci.
Rep. \textbf{5}, 11983 (2015).

\bibitem{16adder}  M. Oszmaniec, A. Grudka, M. Horodecki, and A. W\'{o}jcik, Creating a superposition of unknown quantum states, Phys. Rev. Lett. \textbf{116}, 110403 (2016).

\bibitem{17adder} M. Doosti, F. Kianvash, and V. Karimipour,
Universal superposition of orthogonal states,
Phys. Rev. A \textbf{96}, 052318 (2017).

\bibitem{17adder2} G. Gatti, D. Barberena, M. Sanz, and E. Solano, Protected state transfer via an
approximate quantum adder, Sci.
Rep. \textbf{7}, 6964 (2017).

\bibitem{16huang} X. M. Hu, M. J. Hu, J. S. Chen, B. H. Liu, Y. F. Huang,
C. F. Li, G. C. Guo, and Y. S. Zhang,
Experimental creation of superposition of unknown photonic quantum states, Phys. Rev. A \textbf{94}, 033844 (2016).

\bibitem{17Laflamme} K. Li, G. Long, H. Katiyar, T. Xin, G. Feng, D. Lu, and R. Laflamme,
Experimentally superposing two pure states with partial prior knowledge, Phys. Rev. A \textbf{95}, 022334 (2017).

\bibitem{11you} J. Q. You and F. Nori, Atomic physics and quantum optics using superconducting circuits,
Nature \textbf{474}, 589 (2011).

\bibitem{13Devoret} M. H. Devoret and R. J. Schoelkopf, Superconducting circuits for quantum
information: An outlook, Science \textbf{339}, 1169 (2013).

\bibitem{17liuy} X. Gu, A. F. Kockum, A. Miranowicz, Y. X. Liu, and F. Nori, Microwave photonics with superconducting quantum circuits, Physics Reports \textbf{718}, 1 (2017).

\bibitem{18Romanenko} A. Romanenko, R. Pilipenko, S. Zorzetti, D. Frolov, M. Awida, S. Posen, and A. Grassellino, Three-dimensional superconducting resonators at T $<$ 20 mK with the photon lifetime
up to $\tau = 2$ seconds, arXiv:1810.03703 (2018).

\bibitem{12Rigetti} C. Rigetti, J. M. Gambetta, S. Poletto, B. L. T. Plourde, J. M. Chow, A. D. C\'{o}rcoles, J. A. Smolin,
S. T. Merkel, J. R. Rozen, G. A. Keefe, M. B. Rothwell, M. B. Ketchen, and M. Steffen, Superconducting qubit in a waveguide cavity with a coherence time approaching 0.1 ms, Phys. Rev. B \textbf{86}, 100506(R) (2012).

\bibitem{08Mariantoni} M. Mariantoni, F. Deppe, A. Marx, R. Gross, F. K.
Wilhelm, and E. Solano, Two-resonator circuit quantum electrodynamics: A
superconducting quantum switch, Phys. Rev. B \textbf{78}, 104508 (2008).

\bibitem{njpMerkel} S. T. Merkel and F. K. Wilhelm, Generation and detection
of NOON states in superconducting circuits, New J. Phys. \textbf{12}, 093036
(2010).

\bibitem{10Strauch} F. W. Strauch, K. Jacobs, and R. W. Simmonds, Arbitrary
control of entanglement between two superconducting resonators, Phys. Rev.
Lett. \textbf{105}, 050501 (2010).

\bibitem{15Xiong} S. J. Xiong, Z. Sun, J. M. Liu, T. Liu, and C. P. Yang, Efficient scheme for generation of photonic NOON states in circuit QED, Opt. Lett. \textbf{40}, 2221 (2015).

\bibitem{16Sharma} R. Sharma and F. W. Strauch, Quantum state synthesis of
superconducting resonators, Phys. Rev. A \textbf{93}, 012342 (2016).

\bibitem{16Zhao} Y. J. Zhao, C. Q. Wang, X. B. Zhu, and Y. X. Liu,
Engineering entangled microwave photon states through multiphoton
interactions between two cavity fields and a superconducting qubit, Sci.
Rep. \textbf{6} 23646 (2016).

\bibitem{17su} Q. P. Su, H. H. Zhu, L. Yu, Y. Zhang, S. J. Xiong, J. M. Liu, and C. P. Yang, Generating double NOON
states of photons in circuit QED, Phys. Rev. A  \textbf{95}, 022339 (2017).

\bibitem{18liut} T. Liu, B. Q. Guo, C. S. Yu, and W. N. Zhang, One-step implementation of a hybrid Fredkin gate with quantum memories and single superconducting qubit in circuit QED and its applications, Opt. Express \textbf{26} 4498 (2018).

\bibitem{12yang} C. P. Yang, Q. P. Su, and S. Han, Generation of
Greenberger-Horne-Zeilinger entangled states of photons in multiple cavities
via a superconducting qutrit or an atom through resonant interaction,
Phys. Rev. A \textbf{86}, 022329 (2012).

\bibitem{16liu2} T. Liu, Q. P. Su, S. J. Xiong, J. M. Liu, C. P. Yang, and F. Nori, Generation of a macroscopic
entangled coherent state using quantum memories in circuit QED. Sci. Rep. \textbf{6}, 32004 (2016).

\bibitem{17liu} T. Liu, Y. Zhang, B. Q. Guo, C. S. Yu, and W. N. Zhang,
Circuit QED: cross-Kerr effect induced by a superconducting qutrit without
classical pulses, Quantum Inf. Process. \textbf{16}, 209 (2017).

\bibitem{18deng} H. Zhang, A. Alsaedi, T. Hayat,
and F. G. Deng,
Entanglement concentration and purification of two-mode squeezed microwave photons in circuit QED,
Annals of Physics \textbf{391}, 112 (2018).

\bibitem{s6} H. Paik, D. I. Schuster, L. S. Bishop, G. Kirchmair, G. Catelani, A. P. Sears, B. R. Johnson, M. J. Reagor,
L. Frunzio, L. I. Glazman, S. M. Girvin, M. H. Devoret, and R. J. Schoelkopf, Observation of high coherence in Josephson junction qubits measured
in a three-dimensional circuit QED architecture, Phys. Rev. Lett. \textbf{107}, 240501 (2011).

\bibitem{13Vlastakis} B. Vlastakis, G. Kirchmair, Z. Leghtas, S. E. Nigg, L. Frunzio,
S. M. Girvin, M. Mirrahimi, M. H. Devoret, R. J. Schoelkopf,
Deterministically encoding quantum information using 100-photon Schr\"{o}dinger cat states, 
Science \textbf{342}, 607 (2013).

\bibitem{15transmon} M. J. Peterer, S. J. Bader, X. Jin, F. Yan, A. Kamal, T. J. Gudmundsen, P. J. Leek, T. P. Orlando, W. D. Oliver, and S. Gustavsson, Coherence and decay of higher energy levels of a superconducting transmon qubit, Phys. Rev. Lett. \textbf{114}, 010501 (2015).

\bibitem{16correction} N. Ofek, A. Petrenko, R. Heeres, P. Reinhold, Z. Leghtas, B. Vlastakis, Y. Liu, L. Frunzio, S. M. Girvin, L. Jiang, M. Mirrahimi, M. H. Devoret, and R. J. Schoelkopf, Extending the lifetime of a quantum bit with error correction in superconducting circuits,
Nature \textbf{536}, 441 (2016).

\bibitem{16Wangc} C. Wang, Y. Y. Gao, P. Reinhold, R. W. Heeres, N. Ofek,
K. Chou, C. Axline, M. Reagor, J. Blumoff, K. M. Sliwa,
L. Frunzio, S. M. Girvin, L. Jiang, M. Mirrahimi,
M. H. Devoret, R. J. Schoelkopf, A Schr\"{o}dinger cat living in two boxes, Science \textbf{352}, 1087 (2016).

\bibitem{16Rosenblum} S. Rosenblum, Y. Y. Gao, P. Reinhold, C. Wang, C. J. Axline, L. Frunzio, S. M. Girvin, L. Jiang, M. Mirrahimi, M. H. Devoret, and R. J. Schoelkopf,
A CNOT gate between multiphoton qubits encoded in two cavities, 
Nature Comm. \textbf{9}, 652 (2018).

\bibitem{07dot} K. Hennessy, A. Badolato, M. Winger, D. Gerace, M. Atat\"ure, S. Gulde, S. F\"alt, E. L. Hu, and A. Imamo\u{g}lu, Quantum nature of a strongly coupled single quantum dot-cavity system, Nature \textbf{445}, 896 (2007).

\bibitem{05atom} K. M. Birnbaum, A. Boca, R. Miller, A. D. Boozer, T. E. Northup, and H. J. Kimble, Photon blockade in an optical cavity with one
trapped atom, Nature \textbf{436}, 87 (2005).

\bibitem{11phase} H. Wang, M. Mariantoni, R. C. Bialczak, M. Lenander, E. Lucero, M. Neeley, A. D. O'Connell, D. Sank, M. Weides, J. Wenner, T. Yamamoto, Y. Yin, J. Zhao, J. M. Martinis, and A. N. Cleland, Phys. Rev. Lett. \textbf{106}, 060401 (2011).

\bibitem{13Xmon} R. Barends, J. Kelly, A. Megrant, D. Sank, E. Jeffrey, Y. Chen, Y. Yin, B. Chiaro, J. Mutus, C. Neill, P. O'Malley,
P. Roushan, J. Wenner, T. C. White, A. N. Cleland, and J. M. Martinis, Coherent josephson qubit suitable for scalable quantum integrated circuits, Phys. Rev. Lett. \textbf{111}, 080502 (2013).

\bibitem{16flux} F. Yan, S. Gustavsson, A. Kamal, J. Birenbaum, A. P Sears, D. Hover, T. J. Gudmundsen, D. Rosenberg, G. Samach, S Weber, J. L. Yoder, T. P. Orlando, J. Clarke, A. J. Kerman, and W. D. Oliver, The flux qubit revisited to enhance coherence
and reproducibility, Nature Commun. \textbf{7}, 12964 (2016).

\bibitem{11zhu} X. Zhu,	S. Saito, A. Kemp, K. Kakuyanagi, S. Karimoto, H. Nakano, W. J. Munro, Y. Tokura, M. S. Everitt, K. Nemoto, M. Kasu, N. Mizuochi, and K. Semba, Coherent coupling of a superconducting flux qubit to an electron spin ensemble in diamond, Nature \textbf{478}, 221 (2011).

\bibitem{08M. Neeley} M. Neeley, M. Ansmann, R. C. Bialczak, M. Hofheinz, N. Katz,
E. Lucero, A. O'Connell, H. Wang, A. N. Cleland, and J. M. Martinis, Nat. Physics \textbf{4}, 523 (2008).

\bibitem{09P. J. Leek} P. J. Leek, S. Filipp, P. Maurer, M. Baur, R. Bianchetti, J.
M. Fink, M. G\"oppl, L. Steffen, and A. Wallraff, Phys. Rev. B
\textbf{79}, 180511 (2009).

\bibitem{13J. D. Strand} J. D. Strand, M. Ware, F. Beaudoin, T. A. Ohki, B. R. Johnson,
A. Blais, and B. L. T. Plourde, Phys. Rev. B \textbf{87},
220505 (2013).

\bibitem{07trans} J. Koch, T. M. Yu, J. Gambetta, A. A. Houck, D. I. Schuster, J. Majer, A. Blais, M. H. Devoret, S. M. Girvin, and R. J. Schoelkopf, Charge-insensitive qubit design derived from the Cooper pair box,
Phys. Rev. A  \textbf{76}, 042319 (2007).

\bibitem{07James} D. F. James and J. Jerke, Effective Hamiltonian theory and its applications in quantum information, Can. J. Phys. \textbf{85},
625 (2007).

\bibitem{ini1}  M. D. Reed, L. DiCarlo, B. R. Johnson, L. Sun, D. I. Schuster, L. Frunzio, and R. J. Schoelkopf, High-fidelity readout in circuit quantum electrodynamics using the Jaynes-Cummings nonlinearity, Phys. Rev. Lett. \textbf{105}, 173601 (2010).

\bibitem{ini2} K. S. Chou, Jacob Z. Blumoff, C. S. Wang, P. C. reinhold, C. J. Axline, Y. Y. Gao, L. Frunzio, M. H. Devoret, L. Jiang, and R. J. Schoelkopf, Deterministic teleportation of a quantum gate between two logical qubits, Nature \textbf{561}, 368 (2018).

\bibitem{16sr2} T. Liu, X. Z. Cao, Q. P. Su, S. J. Xiong, and C. P. Yang, Multi-target-qubit unconventional
geometric phase gate in a multi-cavity
system, Sci. Rep. \textbf{6}, 21562 (2016).



\end{thebibliography}
\end{document}